\documentclass[structabstract]{aa}
\pdfoutput=1 
\usepackage{txfonts}
\usepackage{pictex}
\usepackage{latexsym}
\usepackage{graphicx}
\usepackage{afterpage}

\begin{document}


\title
{
Spectral variability of quasars from multi-epoch photometric data in 
the Sloan Digital Sky Survey Stripe 82
}

\author{H. Meusinger      \inst{1}
	 \and 
	 A. Hinze    \inst{2}
         \and 
	 A. de Hoon \inst{3} 
	 }
	 
    \institute{
	 Th\"uringer Landessternwarte Tautenburg, Sternwarte 5, D--07778
         Tautenburg, Germany,
	       e-mail: meus@tls-tautenburg.de
         \and 
 	 Astronomisches Institut,
	 Universit\"at Bern, 
	 Sidlerstra\ss{}e 5, CH--3012 Bern, Switzerland
	 \and
	 Astrophysikalisches Institut Potsdam,
	 An der Sternwarte 16,
	 D--14482 Potsdam, Germany
        }
	
\date{Received / Accepted }

\abstract {}
   {
   The study of the ensemble properties of the UV/optical broadband variability of 
   quasars is hampered by the combined effects of the dependence of variability on 
   timescale, rest-frame wavelength, and luminosity. Here, we present a new approach 
   to analysing the dependence of quasar variability on rest-frame wavelengths. 
   }
   {
  We exploited the spectral archive of the Sloan Digital Sky Survey 
  (SDSS) to create a sample of over 9000 quasars in the Stripe 82. 
  The quasar catalogue was matched with the Light Motion Curve Catalogue for 
  SDSS Stripe 82 and first-order structure functions were computed from the
  lightcurves. The structure functions are used to create a variability indicator 
  that is related to the same intrinsic timescales for all quasars (about 1 to 2\,yr in the
  rest-frame). We study the variability ratios for adjacent SDSS filter bands as a 
  function of redshift. A quantitative interpretation of these relations
  is provided by comparing with the results of simple Monte Carlo simulations
  of variable quasar spectra.  
   }  
   {
   We confirm the well-known dependence of variability on time-lag; the best power-law fit
   of the sample-averaged structure function has a slope $\beta = 0.31\pm0.03$.
   We also confirm that anti-correlations exist with luminosity, wavelength, and redshift, 
	 where the latter can be fully explained as a consequence of the former two dependencies. 
   The variability ratios as a function of redshift resemble the corresponding 
   colour index-redshift relations. While variability is almost always stronger in the 
   bluer passband than in the redder, the variability ratio depends on whether 
   strong emission lines contribute to either one band or the other. 
   We find that the observed variability ratio-redshift relations are described well assuming 
   that (a) the r.m.s. fluctuation of the quasar continuum flux follows a power law
   $\sigma(f_\lambda) \propto \lambda^{-2}$  (i.e., is bluer when 
   brighter) and (b) the variability of the emission line  flux is only $\sim 10$\%
   of that of the underlying continuum. These results, based upon the photometry of 
   more than 8000 quasars, confirm the previous findings by Wilhite and collaborators (2005) 
   for 315 quasars with repeated SDSS spectroscopy. 
   Finally, we find that quasars with unusual spectra and weak emission lines tend to have 
   less variability than conventional quasars. This trend is the opposite of that expected from 
   the dilution effect of variability due to line emission and may be indicative of 
   high Eddington ratios in these unusual quasars.  
     }
   {}
      
\keywords{Galaxies: active --
          Galaxies: statistics --
	  Quasars: general --
	  Quasars: emission lines 
         }

\titlerunning{Spectral variability of quasars in SDSS Stripe 82}
\authorrunning{H. Meusinger et al.}

\maketitle

%
%
\section{Introduction}
%
%

The variation in flux density over time is one of the 
key characteristics of active galactic nuclei (AGN).
From the onset of the study of quasars, which are
the most luminous AGNs, variability is known to be a diagnostic property 
and has been successfully used as a criterion for selecting quasar 
candidates in a number of studies. On the other hand, the physical 
mechanisms behind these fluctuations are still poorly known,
though it has long been understood that the observed flux variations of AGNs 
hold keys to the structure of the radiation source. 
The advent of modern variability surveys (e.g., 
Ivezi\'c et al. \cite{Ivezic03};
Groot et al. \cite{Groot03};
Becker et al. \cite{Becker04}; 
Bauer et al. \cite{Bauer09};
Djorgovski et al. \cite{Djorgovski08};
LSST Science Collaborations \cite{LSST09})
combining high photometric accuracy, large survey areas, sufficiently long 
time baselines, and a high number of observation epochs 
holds the promise of significant progress. 

Numerous studies have shown that the optical/UV continuum variability of quasars 
correlates with various physical parameters such as time-lag, 
rest-frame wavelength, luminosity, radio properties, emission 
line properties, and the Eddington ratio 
(e.g., 
Ulrich et al. \cite{Ulrich97};
Giveon \cite{Giveon99};
Helfand \cite{Helfand01};
Gaskell \& Klimek \cite{Gaskell03};
Vanden Berk et al. \cite{VandenBerk04};
Rengstorf et al. \cite{Rengstorf06};
Wold et al. \cite{Wold07};
Wilhite et al. \cite{Wilhite08};
Bauer et al. \cite{Bauer09};
Ai et al. \cite{Ai10};
Schmidt et al. \cite{Schmidt10}).
Despite considerable progress in the description of quasar variability, 
there are still conflicting claims about these correlations and their
physical interpretation is fraught with problems. 
The most frequently discussed models for the continuum variability of AGNs 
include instabilities in the accretion disk 
(e.g., Kawaguchi et al. \cite{Kawaguchi98}), multiple supernovae
(or other Poissonian processes; e.g., 
Terlevich et al. \cite{Terlevich92};
Cid Fernandes et al. \cite{CidFernandes97}),
star-star collisions in the dense circumnuclear environment 
(Torricelli-Ciamponi et al. \cite{Torricelli00}),
and microlensing by compact foreground objects 
(e.g., Chang \& Refsdahl \cite{Chang79};
Hawkins \cite{Hawkins93}, \cite{Hawkins10};
Lewis \& Irwin \cite{Lewis96},
Zackrisson et al. \cite{Zackrisson03}).

Much of the knowledge about quasar variability is based on the detailed 
investigation of large, statistically well-defined quasar samples.
Even though each individual lightcurve is poorly sampled, a large number of
quasars and the wide range of parameter values allow us to distinguish 
the dependences of variability on the various parameters. 
Hence, substantial improvement has been achieved by employing 
the quasar data base from the Sloan Digital Sky Survey 
(SDSS; York et al. \cite{York00}; Abazajan et al. \cite{Abazajian09}). 
Using a sample of 25\,000 spectroscopically confirmed quasars from the 
SDSS, Vanden Berk et al. (\cite{VandenBerk04}) parametrized correlations 
between variability and time-lag, luminosity, redshift, and rest-frame 
wavelength upon a time-baseline of up to $\sim 2$\,yr in the observer frame. 
A sample of more than 41\,000 quasars was studied by de Vries et al.
(\cite{deVries05}). By combining the SDSS photometric data with 
the Palomar Observatory Sky Survey (POSS), these authors achieved a baseline
of up to 50\,yr in the observer frame, yet with a very sparse lightcurve sampling
per quasar.  

Wilhite et al. (\cite{Wilhite05}) selected a set of 315 quasars with 
multi-epoch SDSS spectroscopy to analyse variability from spectrophotometry. 
In this way, the wavelength dependence of quasar variability could be 
studied at spectral resolution high enough to allow the analysis of variability 
not only in the continuum but also in the major emission lines. The 
results confirm that the continua are bluer when brighter and clearly show 
that variability is weak in the emission lines. They provide
a natural explanation of the intrinsic Baldwin effect behaviour of 
individual AGNs (Kinney et al. \cite{Kinney90}; Wilhite et al. \cite{Wilhite06}).
Pereyra et al. (\cite{Pereyra06}) demonstrated that 
the composite flux differences in the rest-frame wavelength range
1300-6000\,\AA\ can be fit by a standard thermal accretion disk model
in which the accretion rate has changed from epoch to epoch.
These results suggest that the dominant fraction of
the optical/UV variability of quasars is related to the accretion
process. On the other hand, chromatic variability may also be produced by 
quasar microlensing as the  continuum source is compact and bluer 
at smaller radii (Wambsganss \& Paczy\'nski \cite{Wambsganss91};
Yonehara et al. \cite{Yonehara08}, and references therein).
Wilhite et al. (\cite{Wilhite05}) argued that the low 
variability of the broad emission lines provides a problem for
the microlensing scenario as one would 
expect a higher fraction of events where the lense crosses the
broad line region (BLR) but not the compact source. The BLR was once
considered too large to be affected significantly by microlensing. 
Studies have shown, however, that the BLR can be small enough
to be amplified significantly by stellar-mass objects (Abajas et al.
\cite{Abajas07}, and references therein).

A 300 square degree area in the Southern Galactic Cap 
has been repeatedly imaged by the SDSS 
since 1998 and by the SDSS-II Supernova Survey (Frieman et al. \cite{Frieman08}) 
since 2005. The multi-epoch observations in this SDSS Stripe\,82 (S82) 
provide a useful database for the statistical analysis of quasar variability. 
Sesar et al. (\cite{Sesar07}) were the first to analyse the variable 
sources in S82 with the emphasis of characterizing the 
faint variable sky and quantifying the variable population as a whole.
They present methods to select variable sources and discuss their 
distribution in the magnitude-colour-variability space. The three 
dominant classes of variables found by Sesar et al. are quasars (63\%),
RR Lyrae stars (7\%), and stars from the main stellar locus (25\%). 
Moreover, they demonstrate that at least 90\% of the quasars are
significantly variable. An interesting detail of this paper is 
a plot (their figure 7) of the ratio, $V_{\rm g}/V_{\rm r}$, of the 
variability $V$ of quasars in the g- and r-band as a function of redshift $z$. 
Sesar et al. observe a feature in this diagram at $z \sim 1.0 - 1.6$ and argue
qualitatively that it is due different parts of the 
spectrum varying in different ways. This is an important observation 
that offers a new possibility of investigating the spectral variability
of quasars, which is obviously worth studying in detail.

The present paper is focussed on the variability ratios $Q_{\rm k,l} = V_{\rm k}/V_{\rm l}$ 
as a function of redshift. The main aim is to derive the dependence of the variability on 
the (intrinsic) wavelength. In view of the combined effects of the dependence of the 
variability on wavelength, time, and luminosity, the {\it ratios}  
of the variabilities in  different photometric bands provide a better 
approach to the spectral behaviour of variability than variability 
measured in a single band, as was pointed out by Sesar et al. (\cite{Sesar07}). 
The main differences from Sesar et al. are the following:
(i) Sesar et al. used photometric measurements from 58 imaging runs from 1998 
September to 2004 December. In the present study, the lightcurves
are taken from the Light and Motion Curve Catalogue (LMCC) 
of the SDSS S82 (Bramich  et al. \cite{Bramich08}). The catalogue is based on 
134 imaging runs obtained from 1998 to 2005 November with 62 runs obtained in 
2005 alone.
(ii) We extend the approach to all five SDSS bands.
(iii) We use variability indicators related to (more or less) the same
quasar-intrinsic timescale to eliminate $z$-dependent selection effects due
to cosmological time dilation.
(iv) Since the approach works only if quasar variability is dominated by 
the same processes at low and high redshift, we check whether this assumption is 
consistent with our data.
(v) Finally, a quantitative interpretation of the diagrams is presented by means 
of numerical simulations.
Compared to the spectroscopic sample of Wilhite et al. (\cite{Wilhite05}),
the quasar sample of the present study is much larger and the time baseline 
covered by the data is longer. 

The creation of the quasar sample is described in Sect.\,\ref{sample}. Variability 
indicators are computed from their lightcurves in Sect.\,\ref{method}. 
The resulting $Q_{\rm k,l}$-$z$ relations for the five SDSS bands ($k,l = 1\ldots5$)
are presented in Sect.\,\ref{results} and compared to the results of
a simple parametrized model in Sect.\,\ref{interpretation}. 
Other subjects in Sect.\,\ref{interpretation} are the comparison of our results
with Wilhite et al. (\cite{Wilhite05}) followed by a brief discussion of the variability
properties of unusual quasars and the subsample of radio-loud quasars. 
Finally, Sect.\,\ref{conclusion} presents our conclusions. 
Standard cosmological parameters 
$H_0=71$ km s$^{-1}$ Mpc$^{-1}, \Omega_{\rm m}=0.27, \Omega_{\Lambda}=0.73$
are used throughout the paper.

%
%
\section{The quasar sample}\label{sample}
%
%

The SDSS Seventh Data Release (DR7; Abazajan et al. \cite{Abazajian09}), marking
the completion of the original goals of the SDSS and the end of the phase known as 
SDSS-II, contains over 1.6 million spectra in total, including 120\,000 quasars. 
The quasar sample for the present study was constructed from the SDSS DR7 
public archive of spectra in the region of the Stripe 82, i.e., right ascension 
$\alpha = 20^{\rm h}\ldots 4^{\rm h}$ and declination 
$\delta = -1\degr 25 \ldots +1\degr 25$.

Whether a spectrum is useful for scientific purposes or not 
is indicated by the {\sc SciencePrimary} flag, which is set to be either 1 or 0. 
Since {\sc SciencePrimary} = 1 does not necessarily guarantee the 
correctness and precision of the cataloged quantities, we decided
to download all spectra classified as quasars ({\sc Spec\_cln}=3), regardless of their 
SciencePrimary flag, including all spectra classified as unknown
({\sc Spec\_cln}=0). This last decision was motivated (i) by our aim of creating a 
quasar sample as large and complete as possible and (ii) the claim that 
interesting cases of rare unusual quasar spectra (e.g., Hall et al. \cite{Hall02}) 
might be hidden there. The list contains 20\,542 objects of which 19\,660 
were available for download (13\,224 with {\sc Spec\_cln}=3 and 6436 with {\sc Spec\_cln}=0). 
The quasar selection does not include AGNs classified spectroscopically 
as galaxies ({\sc Spec\_cln}=2). We expect the majority of them to be
low-redshift, low-luminosity AGNs, while 
the present study aims to compare the variability at different redshifts
$z=0.2\ldots 3$ and is restricted, therefore, to high-luminosity AGNs ($M_{\rm i} < -21$, 
Sect.\,\ref{sec:var_lambda}). Moreover, the stellar light from the host galaxies 
of low-luminosity AGNs tends to ``dilute'' the variability of the AGN.

All spectra were checked individually to estimate redshifts $z$ and object types.
Redshift values were computed automatically by comparing of measured line 
centres of the prominent emission lines with a catalogue of their rest-frame
wavelengths. The spectral lines useful for the redshift ($z$) estimation were 
selected manually: this allows us to recognize and exclude quasar lines influenced 
by e.g., the remnants of bright night sky lines as well as Ly$\alpha$ lines affected by 
strong absorption of the blue wing. Dubious cases were 
flagged ({\sc Flag\_z} = 1). While the approach 
was not optimized to provide the highest accuracy $z$ values, roughly erroneous
redshift estimations could be efficiently suppressed, which is more important 
for the present study.  The spectra were classified into the six types: 
quasar (q), possible quasar (q?), galaxy (g), star (s), unclear (u), and noise (n). 
For quasar spectra, special features such as strong broad absorption lines (BALs), 
unusual BALs, weak emission lines, or strong iron emission were noted. The 
procedure was performed twice for all 19\,660 spectra.

\begin{table}[hhh]
\caption{Number  of quasars from the basic catalogue with
redshift deviations $|z_{\rm here} - z_{\rm SDSS}| > |\Delta z|_{\rm max}$. 
}
\begin{flushleft}
\begin{tabular}{lrrrrr}
\hline
&&&&&\\
$|\Delta z|_{\rm max} =$      &  0.05  &  0.1   &  0.5  & 1.0 & 2.0 \\ 
&&&&&\\
\hline
&&&&&\\
\multicolumn{6}{l} 8\,212 quasars found in the QC\,DR5: \\
$N (> |\Delta z|_{\rm max}) $  & 22     &  5     &   3   &  1  & 0\\
&&&&&\\
\multicolumn{6}{l} 1\,643 quasars not found in the QC\,DR5: \\
$N (> |\Delta z|_{\rm max}) $  &  391   &  381   & 314   & 167 & 51\\
&&&&&\\
\hline
\end{tabular}\\
\end{flushleft}
\label{tab:z-differences}
\end{table}

We assigned 13\,558 selected spectra to the types q, q?, or g. To distinguish galaxies
from quasars, a simple criterion was used where priority was given to  
low contamination by non-AGNs: spectra containing typical absorption lines and/or 
narrow emission lines but no signs of broad emission-line components were 
classified simply as galaxies. The total number 
of quasar spectra (type q) is 12\,966, among them 976 from the database of the unknowns 
({\sc Spec\_cln}=0), 
another 4 objects were classified as possible quasars. Fiber positions 
that differ by less than the fibre diameter (3 arcsec) can be ascribed with 
high probability to the same object. After excluding multiple observations, 
the resulting basic catalogue contains 9\,855 entries with 2\,021 objects having been 
observed at least two times (660 objects with more than 2 spectra). 
For all objects with multiple spectra, the $z$ and type of the
single spectra were compared with each other. Inconsistent results were found 
in only very few cases (e.g., because of corrupted spectra or low signal-to-noise ratio (S/N)).

\begin{figure}[bhtp]   
\vspace{6.2cm}
\includegraphics[bb=570 20 15 846,scale=0.31,angle=-90,clip]{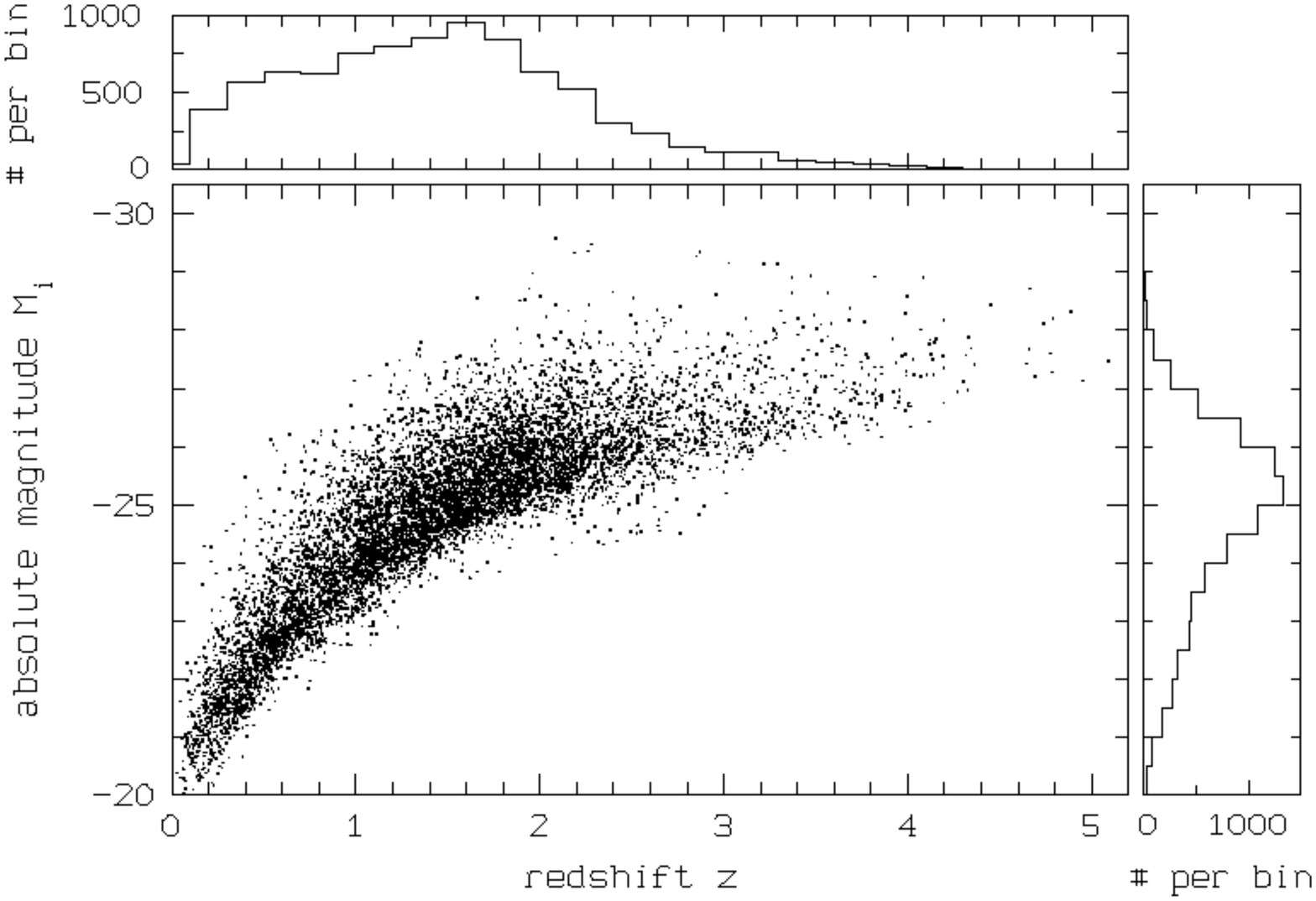}
\caption{
{\it Left:} 
Redshift versus absolute i band magnitude and histograms for the distributions of
redshifts ({\it top}) and absolute magnitudes ({\it right}) for the final quasar sample. 
} 
\label{fig:z-Mi}
\end{figure}

We identified 8\,212 entries in our basic quasar catalogue with entries in
the SDSS DR5 quasar catalogue 
(QC\,DR5; Schneider et al. \cite{Schneider07}), 
736 (7\%) were found only in the database of the unknowns only.
Table\,\ref{tab:z-differences} compares our redshift estimates
with the $z$ values from Schneider et al. (\cite{Schneider07}) 
and the DR7, respectively, for objects not listed 
in the QC\,DR5. The 25 quasars with dubious redshift flags from
the present study were eliminated. While the agreement with the QC\,DR5
is satisfactory, we find that a substantial number of erroneous redshifts 
have been produced by the SDSS DR7 pipeline, which are attributed to erroneous
line identifications (e.g., \ion{Mg}{II} with Ly$\alpha$, 
\ion{C}{IV}, or \ion{C}{III}]) in most cases.

To analyse variability, the basic quasar catalogue was matched to
the LMCC (Bramich et al. \cite{Bramich08}) using an identification radius of 2\,arcsec. 
The LMCC was trimmed to only those objects that have a mean right ascension in the 
range $\alpha = 20.7^{\rm h}\ldots 3.3^{\rm h}$ because of the sparse temporal coverage
outside these limits. Compared to the whole S82, the area of sky covered by the 
LMCC is thus reduced by $\sim 17$\% to $\sim 249$\,deg$^2$. The final quasar catalog 
contains 8\,744 objects identified in the LMCC (Fig.\,\ref{fig:z-Mi}). 
The redshifts cover the range 
$z \approx 0.1 \ldots 5$ with $\overline{z}=1.46\pm0.79$.
Absolute magnitudes $M_{\rm i}$ were computed following Kennefick and Bursick 
(\cite{Kennefick08}; their equations 1 to 6) correcting the SDSS i band 
magnitude for Galactic extinction and applying a standard continuum  $K$ correction 
by assuming a spectral index $\alpha_\nu = -0.5$. The Galactic foreground extinction was
obtained from the ``Galactic Dust Extinction
Service''\footnote{http://irsa.ipac.caltech.edu/applications/DUST/} 
of the NASA/IPAC Infrared Science Archive using the translation relations for the
SDSS bands given by Schlegel et al. (\cite{Schlegel98}).
The mean value of the absolute magnitudes in the sample is 
$M_{\rm i} = -24.7\pm1.57$. The distribution of the quasars in the 
$z$-$M_{\rm i}$ plane is shown in Fig.\,\ref{fig:z-Mi}.

\begin{figure}[bhtp]   
\includegraphics[bb=25 55 570 790,scale=0.45,angle=0,clip]{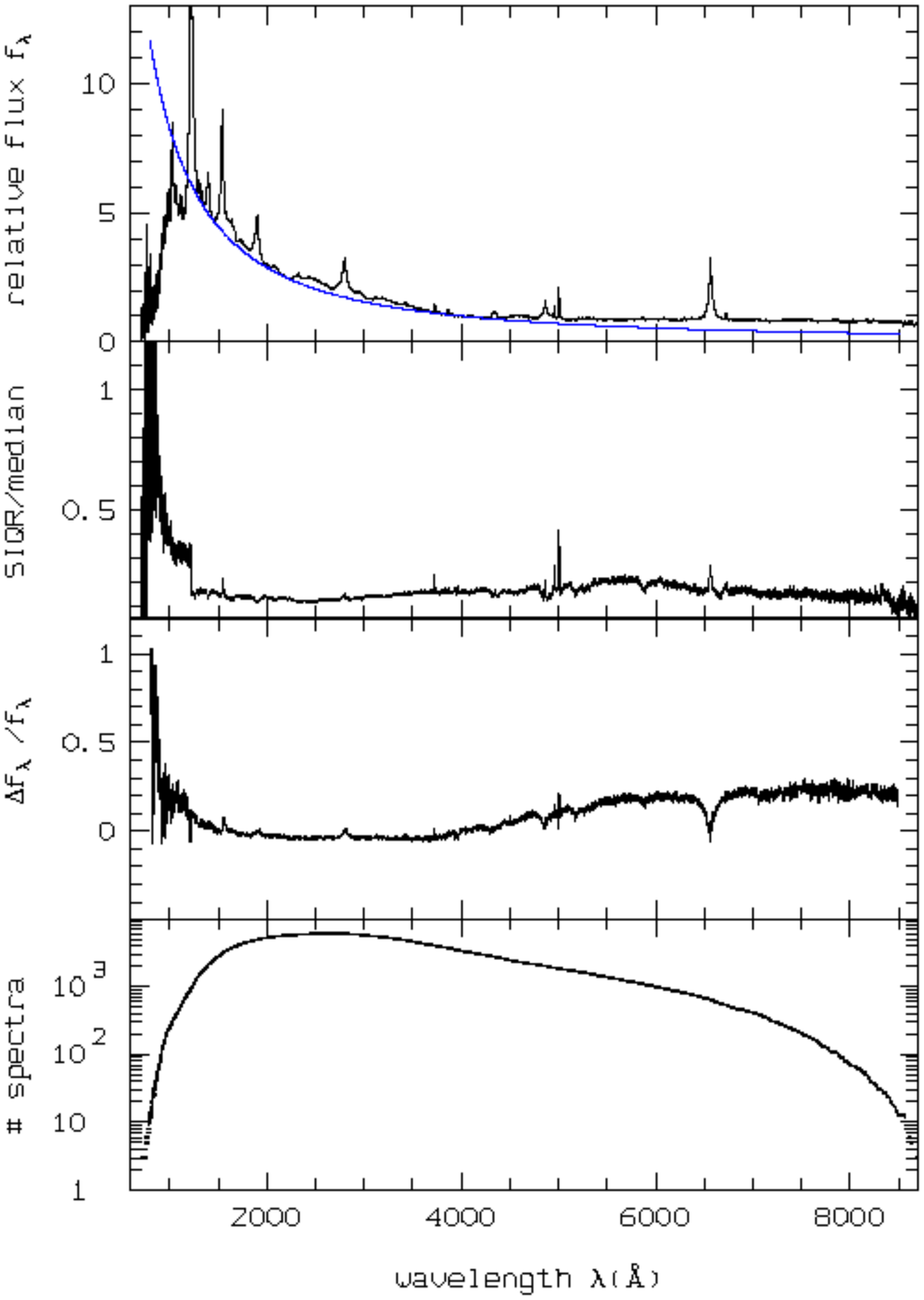}
\caption{
{\it Top:} Median composite spectrum from the final quasar sample with fitting
power-law continuum.
{\it Second row:} The semi-interquartile range, SIQR, in units of the median. 
{\it Third row:} Relative deviation from the Vanden Berk et al. (\cite{VandenBerk01}) 
composite ([here-Vanden Berk]/here). 
{\it Bottom:} Number of spectra per wavelength bin used for the 
construction of the composite. 
} 
\label{fig:composite}
\end{figure}

To model spectral variability (Sect.\,\ref{sim}), we needed to produce 
an average quasar spectrum.  We used our ``standard sample'' 
(see Sect.\,\ref{data} below) 
and constructed a composite following 
Vanden Berk et al. (\cite{VandenBerk01}). As we are primarily 
interested in ``normal'' quasars, objects with highly unusual spectra 
(remark = 'x\_bal' or 'myst' in Table\,\ref{tab:variabilities}) 
are not included as well as quasars with uncertain redshifts. 
We also note that the sample is 
corrected for multiple observations; quasars with two or more SDSS spectra 
are thus represented by only one spectrum, usually the one with the highest
S/N. All spectra were resized to a common range of 3900-9100\AA\ in the 
observer frame, sorted by increasing redshift, corrected for Galactic 
extinction, shifted into their rest-frame, and 
binned to a resolution of 1\AA\ per pixel. 
Subsequently each spectrum was inserted as a new row into a two-dimensional 
image, with 1\AA\ binning on the horizontal axis and one spectrum per pixel 
on the vertical axis. Every time before a new spectrum was inserted,
a preliminary mean quasar spectrum was created simply by computing  
the mean pixel value for each wavelength bin in the two-dimensional 
image. The result is used for the normalisation: before inserting into 
the two-dimensional image, each spectrum is normalised to the mean pixel
value of the overlapping region with the respective preliminary composite 
spectrum. The spectra was not assigned any statistical weights, i.e. the 
S/N of the spectra is assumed to be relatively constant 
throughout the sample and over the observed window. Pixels flagged 'bad' 
in the SDSS 2D spectroscopic pipeline are interpolated over, 
and outliers and negative fluxes were excluded.

Composite spectra can be 
created in different ways. Vanden Berk et al. (\cite{VandenBerk01}) 
discussed how arithmetic and geometric mean composites have different 
properties. Following the arguments given by Wilhite et al. (\cite{Wilhite05}),
we consider the arithmetic composite but we prefer to use the median 
instead of the mean. The final composite spectrum is constructed by computing the 
median value for all wavelength bins represented by at least three spectra. 
The result is shown in the top panel of Fig.\,\ref{fig:composite}. 

Vanden Berk et al. (\cite{VandenBerk01}) suggested using
the semi-interquartile range, SIQR$=(q_3-q_1)/2$ as a measure of the 
spectrum-to-spectrum variation, where $q_1$ and $q_3$ are the first (25\%) 
and third (75\%) interquartile; the second interquartile, $q_2$, is the median.
This ratio, SIQR$/q_2$, is shown in the second panel of Fig.\,\ref{fig:composite}. 
There exist many spectra with a relative flux much higher or lower 
than the median at wavelengths where SIQR$/q_2$ is high .
On the other hand, if SIQR$/q_2$ is small, the distribution is 
sharply peaked. As expected, the variation is huge in the wavelength range
of the Ly$\alpha$ forest. At longer wavelengths, the variations are strong
in the peaks of the emission lines, in particular of the [\ion{O}{iii}] lines.
Longward of $\sim 4000$\AA\, the spectrum-to-spectrum variation 
increases because of the different strong contribution of the host galaxies.

The third panel of Fig.\,\ref{fig:composite} compares our median composite
with that of Vanden Berk et al. (\cite{VandenBerk01}); 
where both spectra have been normalized normalized at 4000\,\AA.\ For wavelengths 
$\lambda \sim 1000\ldots 4000$ \AA, our result 
(Fig.\,\ref{fig:composite}) is in accord with the Vanden Berk composite.  
The deviation at longer wavelengths reflects the 
differences in the quasar selection for the two samples.
Vanden Berk et al. define ``quasar to mean any extragalactic object with 
at least one broad emission line and that is dominated by a nonstellar 
continuum''. Spectra with continua dominated by stellar features were 
rejected by these authors ``to avoid introducing a significant spectral
component from the host galaxies''. This criterion was not applied in 
the present study. The consequence is a stronger contribution from host 
galaxies in our spectrum, yet the absolute magnitude-redshift plot 
appears to have a similar representation from low-luminosity AGNs in both
cases. At short wavelengths, the differences are likely statistical
fluctuations due to the small number of spectra included (bottom panel) 
in combination with the strong spectrum-to-spectrum variations at these
wavelengths.
In the spectral region between Ly\,$\alpha$ and 4000\,\AA, the continuum 
is matched by a power law $f_\lambda \propto \lambda^{\alpha_\lambda}$ 
with $\alpha_\lambda = -1.52$ ($\alpha_\nu = -0.48$, in good agreement
with $\alpha_\nu=-0.46$ given by Vanden Berk et al. for their arithmetic 
median composite).

The quasar catalog was matched to the FIRST (Becker et al. \cite{Becker95}) 
VLA 20\,cm catalogue (match radius 2 arcsec). For the 467 identified FIRST
sources, the values for the integrated 20\,cm flux and the peak flux were 
entered. A radio compactness parameter and the radio loudness parameter 
$R_{\rm i} = \mbox{log}\,(F_{\rm radio}/F_{\rm i})$ 
were computed following Ivezi\'c et al. (\cite{Ivezic02}); 381 sources are radio-loud
($R_{\rm i} > 1$).  Finally, more than 50 quasars were marked as 
having more or less unusual spectra, i.e., 
spectra that deviate remarkably from the typical quasar spectrum. 
The classification was performed by eye and is thus not based on quantifiable
properties. The spectra of these objects display various unusual features 
that are difficult to discern at first glance, often (but not exclusively)
related to complex broad absorption line (BAL) troughs combined with strong reddening 
(e.g., Hall et al.\cite{Hall02}). This subsample is inhomogeneous and includes
unusual BAL quasars as well as weak emission line quasars (WLQs) and
``mysterious'' objects such as \object{SDSS\,J010549.75-003314.0} and 
\object{SDSS\,J220445.26+003142.0} from Hall et al. (\cite{Hall02}). 
In several cases, the redshifts are uncertain. For one of these
objects, \object{SDSS\,J014349.14+002128.4}, the spectrum was first classified as 
an unusual BAL quasar but later rejected after we found it to be
a superimposition of the spectra of a normal quasar in the blue and an
M star in the red.

%
%
\section{Measuring spectral variability}\label{method}
%
%

\subsection{Photometric data}\label{data}

The catalogue of objects in S82 compiled by Bramich et al. (\cite{Bramich08}) contains 
about 3.7 million stellar objects and galaxies. It comes in two flavours, the Light-Motion Curve 
Catalogue (LMCC), which contains the measured quantities for each object listed as a function
of filter band and epoch, and the Higher-Level Catalogue (HLC), which 
contains similar quantities for each light-motion curve. Since the present 
paper intends to analyze the lightcurves, the LMCC is used here. 
For the overwhelming majority ($\ga 90$\,\%) of the quasars, the number of measurement epochs 
is between 20 and 50 with a mean at $\sim35$ in all five bands.  
For a detailed description of the LMCC, we refer to Bramich et al. 
(\cite{Bramich08}) and address only three issues: 
(i) for an object record to be included in the LMCC, a list of constraints 
need to be satisfied in at least one waveband. In addition, we accept here
only measurements, in each band, with the quality flags  
{\sc BRIGHT}=0, {\sc EDGE}=0, {\sc BLENDED}=0, and {\sc SATUR}=0. 
(ii) As noted by Bramich et al., the catalogue is contaminated by some strong 
photometric outliers that are clearly due to `bad epochs'. We find that
some lightcurves show one, sometimes two, rarely more, examples of such 
low quality data that can significantly affect the measurement of variability. We applied a 
2.5$\sigma_{100}$-clip criterion to identify and eliminate the corresponding data 
points, where $\sigma_{100}$ is the standard deviation of magnitudes in a 100\,d time interval
centred on the corresponding epoch. Each lightcurve was checked for up to 
three of these outliers.     
(iii) We restrict the analysis, in each band, to objects brighter than the 
limiting magnitudes 21.9,23.2,22.8,22.2, and 20.7 for {\it u,g,r,i,} and {\it z}. 
More precisely, only quasars with mean magnitudes 0.5 mag brighter than the 
corresponding limit in each band were considered to minimize a 
possible bias by a one-sided censorship of the lightcurves due to variability.

As a first step, we estimate the fraction of significantly variable quasars.
We use the $\chi^2_{1;\alpha}$ per degree of freedom defined by Sesar et al. 
(\cite{Sesar07})
to compare the magnitude fluctuations of the quasars from the final sample with
those of 1\,100 objects selected from the SDSS standard star catalogue for S82 (Iveci\'z et al.
\cite{Ivezic07}). The constraint $\chi^2_{1;0.05} > 3$ (significance level $\alpha=0.05$)
is a useful criterion for significant quasar variability, i.e., the 
fluctuations cannot be explained by Gaussian-distributed measurement 
errors with the same (magnitude-dependent) standard deviations as 
measured for the standard stars. Using this criterion, we find that 93\%, 97\%, 93\%, 
87\%, and 37\% of the quasars are variable in the u,g,r,i, and z band, 
respectively.

\begin{figure}[bhtp]   
\vspace{5.8cm}
\includegraphics[bb=565 40 30 815,scale=0.32,angle=-90,clip]{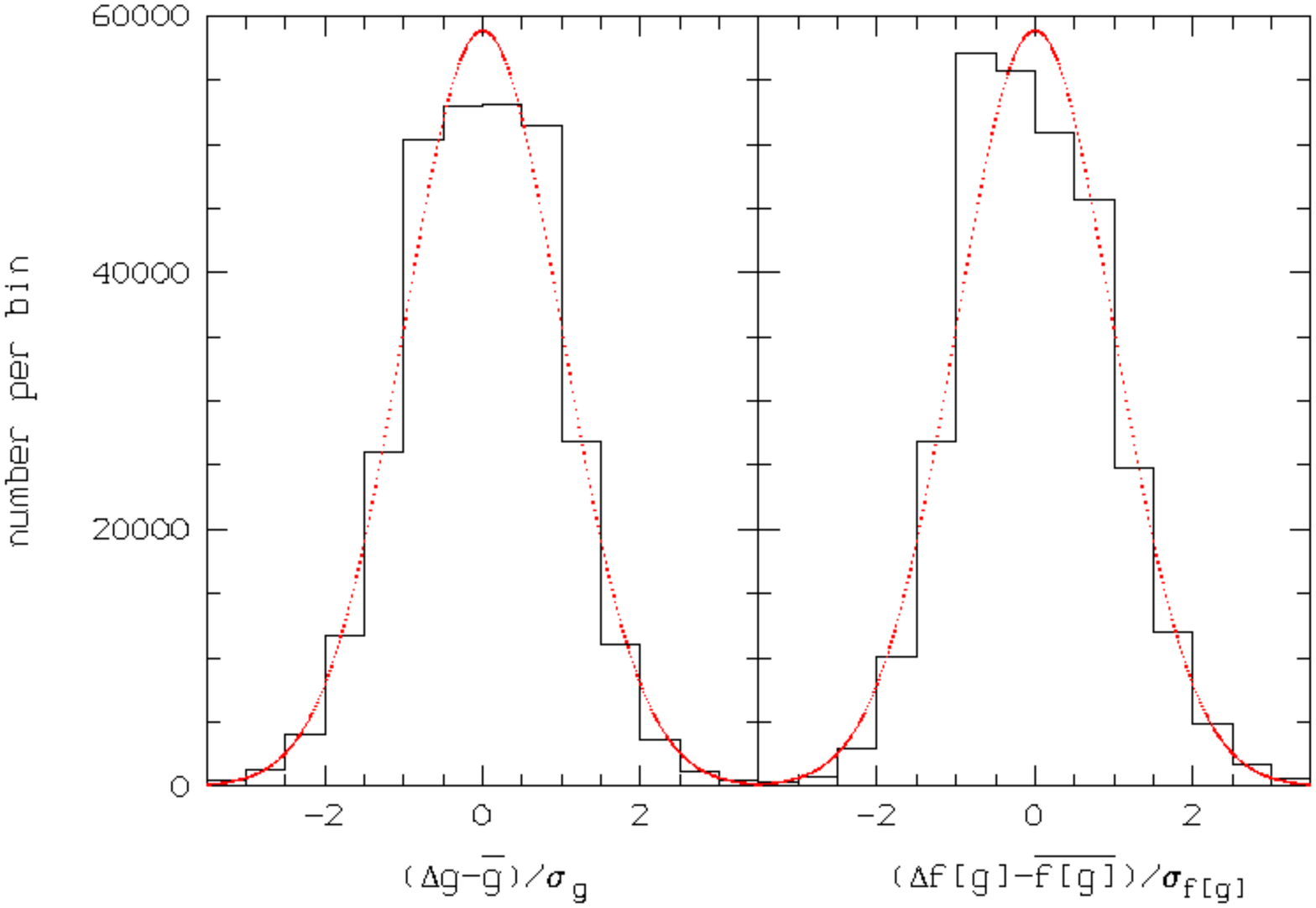}
\caption{
Distribution of the normalized fluctuations of g-band magnitudes ({\it left}) and fluxes 
({\it right}), respectively. Dotted curves: normalized, centred Gaussian distribution. 
} 
\label{fig:gauss}
\end{figure}

We next consider whether variability is characterized more accurately by a normal
distribution of fluxes or magnitudes. Fig.\,\ref{fig:gauss} shows the distribution of the 
normalized fluctuations $\Delta x_{\rm i,j} = (x_{\rm i,j}-\bar{x}_{\rm j})/\sigma_{\rm x;j}$
for the g-band measurements $i$ of the significantly variable quasars $j$, where $x$ 
represents either magnitudes (left hand side) or fluxes (right hand side) after 
the exclusion of outliers. The data are obviously more accurately matched in the
left panel, yet the agreement is not perfect. 
found for both the optical and X-ray radiation of AGNs 
(Gaskell \& Klimek \cite{Gaskell03}, and references therein).

As we are interested 
in the variability properties of typical, ``normal'' quasars, 
the unusual BAL quasars and the quasars with uncertain redshifts were removed 
from the sample. We also rejected  the core-dominated radio-loud quasars. 
Vanden Berk et al. (\cite{VandenBerk04}) and Rengstorf et al. (\cite{Rengstorf06})
found the median integrated radio-flux density to be higher for the variable
population of their samples than for the nonvariable population and suggested 
blazars as a possible explanation. The continuum emission mechanism for
blazars differs fundamentally from that of other quasars (e.g., Ulrich et al.
\cite{Ulrich97}). It is 
unlikely that all rejected radio-loud quasars belong to the blazar 
family but there might be a substantial fraction where both the flux and its variation 
are dominated by the relativistic outflow. In the following, we consider
as our ``standard sample'' all quasars with known redshifts from the final sample
without the unusual and the known radio-loud quasars. The latter two subsamples are 
discussed separately in Sect.\,\ref{unusuals}.

\begin{table*}
\caption{
First five rows of the table containing the variability indicators $V_{\rm k}$. The
full table is available at the CDS.
}
\begin{tabular}{rrrrrrrrrrrrr}
\hline
&&&&&&&&&&&&\\
number & RA       &     DEC   &  redshift &  flag   &  $g$  & $M_{\rm i}$ & $V_{\rm u}$ & $V_{\rm g}$ & $V_{\rm r}$ & $V_{\rm i}$ & $V_{\rm z}$  & remark \\ 
       & (degrees)& (degrees) &           &         &       &             & (mag$^2$)   & (mag$^2$)   &  (mag$^2$)  &  (mag$^2$)  &  (mag$^2$)   &        \\ 
&&&&&&&&&&&&\\
\hline
&&&&&&&&&&&&\\
   1 &   0.001997 & $-0.451102$ & 0.250 & 0 & 21.257 & $-20.39$ & 0.000 & 0.034 & 0.012 & 0.013 & 0.004 &    \\     
   2 &   0.027231 & $ 0.515332$ & 1.823 & 0 & 20.212 & $-25.20$ & 0.073 & 0.122 & 0.088 & 0.027 & 0.009 & 	 \\
   3 &   0.033946 & $ 0.276292$ & 1.837 & 0 & 20.213 & $-25.75$ & 0.012 & 0.025 & 0.017 & 0.005 & 0.013 & 	 \\
   4 &   0.049839 & $ 0.040359$ & 0.479 & 0 & 17.859 & $-24.15$ & 0.044 & 0.017 & 0.005 & 0.003 & 0.000 & 	 \\
   5 &   0.051084 & $-0.539051$ & 1.436 & 0 & 20.435 & $-24.56$ & 0.179 & 0.165 & 0.091 & 0.057 & 0.080 & 	 \\
\hline
\end{tabular}
\label{tab:variabilities}
\end{table*}

\subsection{Structure functions}\label{sf}

The statistical measure of quasar variability used in the present paper is based on 
the first-order structure function (SF) $D(\tau)$, where $\tau$ is the time-lag 
between two observations. The SF is closely related to the autocorrelation function 
and contains, as a sort of running variance (as a function of the time-lag),
information about the timescales of the involved variability processes. 
The SF was first introduced to quasar research in radio astronomy  
(Simonetti et al. \cite{Simonetti85};  
Hjellming and Narayan \cite{Hjellming86};
Hughes et al. \cite{Hughes92})
and has also become a popular tool in optical studies of both quasar samples and 
individual quasar lightcurves 
(e.g., Giallongo et al. \cite{Giallongo91};
Hook et al. \cite{Hook94}; 
Meusinger et al. \cite{Meusinger94};
Kawaguchi et al. \cite{Kawaguchi98}; 
Vanden Berk \cite{VandenBerk04};
De Vries et al. \cite{deVries05};
Wilhite et al. \cite{Wilhite08};
Bauer et al. \cite{Bauer09};
Schmidt et al. \cite{Schmidt10}).
Scholz et al. (\cite{Scholz97}) were the first to use 
an index for long-term variability based on the observed SF to identify 
quasar candidates in an variability-based quasar survey 
(see also Meusinger et al. \cite{Meusinger02}).

There are three different definitions of the SF used in the literature. Here, we
follow the version introduced in the pioneering papers of Simonetti et al. 
(\cite{Simonetti85}) and Hjellming and Narajan (\cite{Hjellming86}), where 
the first-order SF is defined as 
\begin{equation}
D(\tau) = \langle [s(t+\tau)-s(t)]^2 \rangle_{\rm t},
\label{eqn:SF}
\end{equation}
where $s(t)$ is the measured signal at time $t$, and the angular brackets denote the 
time-average. The SF analysis in the optical/UV usually identify signals with
apparent magnitudes, i.e., $s(t) = m(t)$. In a second approach (e.g., 
Di\,Clemente et al. \cite{DiClemente96};
Vanden Berk et al. \cite{VandenBerk04};
Wilhite et al. \cite{Wilhite08}), 
$\langle [\Delta m]^2 \rangle$ 
is replaced by  $\pi/2\,\langle |\Delta m| \rangle^2$.
The former was referred to as SF$^{\rm (A)}$ and the 
latter as SF$^{\rm (B)}$  by Bauer et al. (\cite{Bauer09}). 
In a third version, here called SF$^{\rm (C)}$, 
$\langle |\Delta m| \rangle$ was used instead of 
$\langle [\Delta m]^2 \rangle$ where $\langle\rangle$ means
either the median (Hook et al. \cite{Hook94}) or the mean
(Schmidt et al. \cite{Schmidt10}).   
SF$^{\rm (A)}$ has the advantage of being directly related to other 
statistical quantities such as the variance and the autocorrelation function,
while SF$^{\rm (B)}$ and SF$^{\rm (C)}$ are less sensitive to the presence of outliers in the data.
Since we rejected the outliers in the process of constructing the lightcurves 
(Sect.\,\ref{data}), we decided to apply the ``classical'' definition from Eq.\,\ref{eqn:SF}.

The general definition and some properties of the SF 
are given by Simonetti et al. (\cite{Simonetti85}), an ``ideal'' SF of a ``typical''
measured process is schematically discussed by Hughes et al. (\cite{Hughes92}).
For lags shorter than the smallest correlation timescale, $T_0$, 
of the variable process, the SF is given by $D_0 \equiv D(\tau\ll T_0) = 2\xi^2$, 
where $\xi^2$ is the variance of the measurement noise. 
The SF corrected for the measurement noise is given by 
$D_{\rm corr}(\tau) = D(\tau) -  2\xi^2$. 
For $\tau \gg T_1$, the SF displays another plateau
with $D(\tau) = 2\sigma^2$, where $T_1$ is the longest correlation timescale and 
$\sigma^2>\xi^2$ is the variance of the process. For the intermediate part, 
$T_0 \la \tau \la T_1$, the SF of a stationary random process is characterized by 
a power law $D(\tau) \propto \tau^x$, where the exponent depends on the power 
spectrum of the process.      

We briefly note two other fundamental properties of the SF.
First, the method is suited to the analysis of sparsely 
sampled lightcurves of non-periodic and non-sinusoidal processes. Second, provided 
that variability is quasar-intrinsic, the data obtained at the same epochs for 
quasars at different $z$ cover different source-intrinsic timescales because of the 
cosmological time-dilation and are thus not directly comparable. This problem can be
avoided using the SF as a function of the rest-frame time-lag 
$\tau_{\rm r} = \tau_{\rm o}/(1+z)$. 
Giallongo et al. (\cite{Giallongo91}) were the first to introduce an indicator
for optical quasar variability based on the rest-frame SF. 

In the following, variability is measured by the first-order 
structure function, corrected for measurement errors and binned into rest-frame
time-lag intervals   
\begin{equation}
V = \langle D(\tau_{\rm r}) \rangle_{\tau_{\rm r}}-2 \xi^2(\overline{m}),
\end{equation}\label{eqn:V} 
where $\xi^2(\overline{m})$ is the variance in the 
measurement noise for starlike objects of the apparent magnitude $m=\overline{m}$ 
adopted from the standard stars, where $\overline{m}$ is the mean magnitude of the quasar.
The angular brackets denote the arithmetic mean over a suitable interval of 
rest-frame time-lags around $\tau_{\rm r}$.

\subsection{Spectral variability}\label{spec_var}

For the investigation of the wavelength dependence, the variability
is measured by one single value $V_{\rm k}$ for each quasar in each 
of the five SDSS bands $k = 1 \ldots 5$.  
Distinguishing the $V(\lambda,L,z)$ relations requires two steps. As suggested 
by Sesar et al. (\cite{Sesar07}), we consider the ratio  
$Q_{\rm k,l} \equiv V_{\rm k}/V_{\rm l}$
of the variabilities in two adjacent photometric bands ($k = 1\ldots4, l = 2\ldots5$) 
instead of $V_{\rm k}$ itself. Second, to eliminate $z$-dependent selection effects due to 
cosmological time dilation, the $\tau_{\rm r}$ interval has to be chosen 
properly. The latter is constrained by the following criteria: 
\vspace{-0.2cm}
\begin{itemize}
\item
Equation\,\ref{eqn:SF} yields discrete data points for each quasar
in the $D$-$\tau$ diagram. The mean SF $\langle D(\tau_{\rm r}) \rangle_{\tau_{\rm r}}$
is computed by averaging these data within discrete $\tau_{\rm r}$ bins. 
The binning intervals must be wide enough to cover a large number of 
data points. 
\item
$D(\tau_{\rm r})$ must be related to the same $\tau_{\rm r}$ interval for all quasars
independent of $z$. 
\item
Because quasar variability increases monotonically with $\tau_r$ up
to timescales of $\sim 40$\,yr (deVries et al. \cite{deVries05}), $\tau_r$ 
should be as long as possible.
\item
The maximum rest-frame time-lag is $\tau_{\rm r, max} \sim 7$\,yr (for $z \sim 0$), 
set by the time baseline of the LMCC.
\item
To investigate the wavelength dependence, variability has to
be measured in narrow $z$ bins (see below). The $z$ distribution for our final 
quasar sample (Fig.\,\ref{fig:z-Mi}) shows a strong decline at $z \ga 2$. 
Beyond $z \sim 3$, the number of quasars per $z$ bin becomes too small for a 
statistical analysis. 
\end{itemize}
\vspace{-0.2cm}
Taken together, we find that
$\tau_{\rm r} = 300 - 600$\,days is an appropriate binning interval. 
We computed $\langle D(\tau_{\rm r}) \rangle_{\tau_{\rm r}}$ to be 
the arithmetic mean of all $D(\tau_{\rm r})$ in this $\tau_{\rm r}$ interval.
The value  $V_{\rm k}(\tau_{\rm r})$ from the SF binned in this way 
is considered a useful variability indicator.  
The present analysis is thus related to variability processes with typical 
timescales of the order of $\sim 1 \ldots 2$\ years (rest-frame).  
The median rest-frame time-lag
of 430\,d is a factor of 3.5 longer than for the spectroscopic sample of
Wilhite et al. (\cite{Wilhite05}).  This is an important advantage of the present data    
because of the increase in variability with time-lag (Fig.\,\ref{fig:sf}).

The variability indicators $V_{\rm k}$ for the 8\,744 quasars from our final sample 
are given in Table\,\ref{tab:variabilities}, available at the CDS. The table contains
the following information. Column\,1 is the runnung number, Columns\,2 and 3 list 
right ascension and declination (J\,2000) taken from 
the fits headers of the SDSS spectra. Columns\,4 and 5 give the redshift and the
redshift flag (=1 for uncertain redshifts, 0 otherwise). The g band magnitude
is given in Col.\,6 and the absolute magnitude $M_{\rm i}$ is given in 
Col.\,7. The following five columns contain the variabilities $V$ for ugriz.
For $\langle D(\tau)\rangle < 2\xi^2$, the value of $V_{\rm k}$ was set 
to zero, because negative values of $V_{\rm k}$ are unphysical.
The last column contains information on specific spectral features 
(w\_l: weak-line quasar, s\_bal: strong BAL structure, x\_bal: extreme BAL structure,
myst: mysterious spectrum).

%
%
\section{Results}\label{results}
%
%

\subsection{Variability as function of time-lag}

\begin{figure}[bhtp]   
\vspace{6.5cm}
\includegraphics[bb=580 1 10 840,scale=0.31,angle=-90,clip]{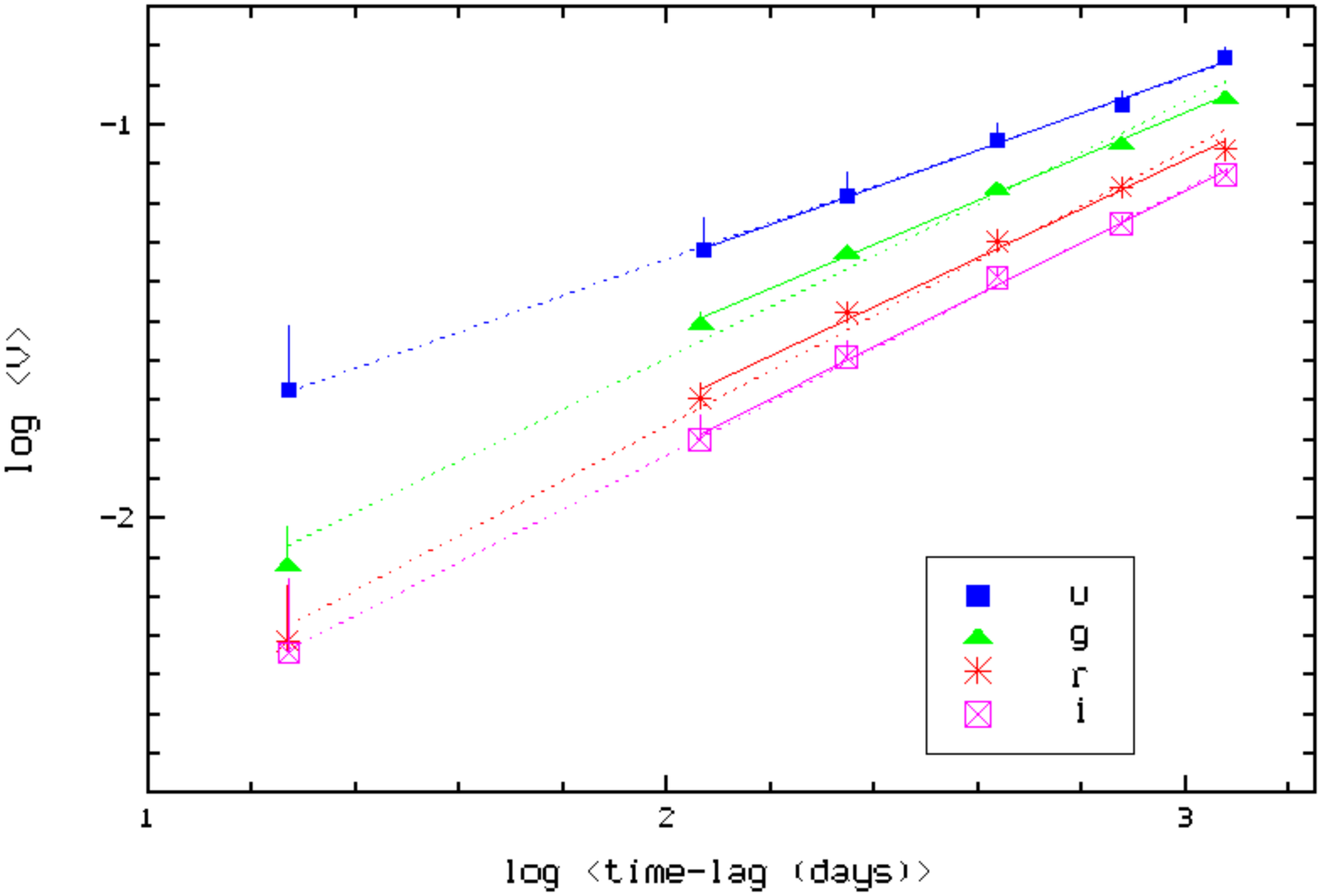}
\caption{
Corrected ensemble-averaged structure functions for the quasars
with $M_{\rm i} < -22$ in the u,g,r, and i 
band as a function of time-lag. The time-lags are in the rest-frame, yet the 
SDSS bands are in the observer frame.
} 
\label{fig:sf}
\end{figure}

The binned, ``noiseless'' first-order structure functions averaged over 
our standard sample of quasars with $M_{\rm i} < -22$ for u,g,r,i, 
respectively, are shown in Fig.\,\ref{fig:sf}. 
Seven time-lag bins were chosen in such a way that we have, on the one hand, 
log $\tau_0$ intervals of comparable widths and, on the other,  reasonably 
large numbers of measurements in each bin. The vertical bars indicate the 
shift due to the corrections for instrumental errors ,which are large on short 
time-lags.
As is evident from Fig.\,\ref{fig:sf}, variability increases with time-lag and 
decreases with rest-frame wavelength. The detailed investigation of 
the dependence of variability on rest-frame wavelength is the subject 
of Sect.\,\ref{sec:var_lambda}. Here we discuss the dependence on time-lag.

\begin{table}[hhh]
\caption{SF exponent $\beta$ for different parameter selections. 
}
\begin{flushleft}
\begin{tabular}{rrrr}
\noalign{\smallskip}
\hline
\noalign{\smallskip}
&&&\\
                              & $M_{\rm i} <-21$& $M_{\rm i} <-22$& $M_{\rm i} <-23$\\                          & \\
			      
\noalign{\smallskip}
\hline
&&&\\
$\tau_{\rm r} > 10$\,days\hspace{0.3cm} 
                            u & 0.230$\pm0.011$ & 0.231$\pm0.010$ & 0.231$\pm0.011$\\
                            g & 0.314$\pm0.031$ & 0.326$\pm0.038$ & 0.331$\pm0.046$\\
                            r & 0.322$\pm0.024$ & 0.326$\pm0.034$ & 0.368$\pm0.048$\\
                            i & 0.305$\pm0.010$ & 0.339$\pm0.013$ & 0.373$\pm0.025$\\
&&&\\
\noalign{\smallskip}
\hline
&&&\\			    
$\tau_{\rm r}> 100$\,days\hspace{0.2cm}   
                            u & 0.237$\pm0.010$ & 0.236$\pm0.009$ & 0.238$\pm0.011$\\
                            g & 0.278$\pm0.010$ & 0.279$\pm0.010$ & 0.272$\pm0.009$\\
                            r & 0.299$\pm0.017$ & 0.312$\pm0.019$ & 0.322$\pm0.021$\\
                            i & 0.311$\pm0.009$ & 0.331$\pm0.012$ & 0.345$\pm0.013$\\
&&&\\			    
\hline
\end{tabular}\\
\end{flushleft}
\label{tab:sf}
\end{table}

\begin{table}[hhh]
\caption{SF exponent $\beta$ from previous work and the present study. 
}
\begin{flushleft}
\begin{tabular}{lcrc}
\noalign{\smallskip}
\hline
&&&\\
Reference                               & $\beta$       &  $\tau_{\rm max}$ (yr) & SF\\
&&&\\
\hline
&&&\\
Hook et al. (\cite{Hook94})             & $0.20$        &   6 (rf)          & C\\
Kawaguchi et al. (\cite{Kawaguchi98})   & $0.35$       	&   2 (of)          & A\\
Hawkins (\cite{Hawkins02}), Quasars     & $0.20\pm0.01$ &  20 (of)          & A\\
Hawkins (\cite{Hawkins02}), Seyferts    & $0.36\pm0.02$ &  20 (of)          & A\\
Vanden Berk et al. (\cite{VandenBerk04})& $0.25\pm0.01$ &   2 (rf)          & B\\
deVries et al. (\cite{deVries05})       & $0.30\pm0.01$ &  50 (of)          & A\\
Rengstorf et al. (\cite{Rengstorf06})   & $0.47\pm0.07$ &   2 (rf)          & B\\
Wilhite et al. (\cite{Wilhite08})       & $0.46\pm0.03$ &   2 (rf)          & B\\
Bauer et al. (\cite{Bauer09})           & $0.36\pm0.01$ &   3 (rf)          & A\\
Schmidt et al. (\cite{Schmidt10})       & $0.43\pm0.002$&  10 (of)          & C\\
&&&\\
present study                           & $0.31\pm0.03$ &   3 (rf)          & A\\
&&&\\			    
\hline
\end{tabular}\\
\end{flushleft}
\label{tab:sf_lit}
\end{table}

In each of the four bands, the SF is closely fitted by a power law 
$[V(\tau_{\rm r})]^{1/2} \propto \tau_{\rm r}^{\ \beta}$ 
(following the notation of Kawaguchi et al. \cite{Kawaguchi98}). 
No significant flattening of the SF is observed at long time-lags.
The slope of $V(\tau)$ in the double-logarithmic presentation depends slightly
on the chosen lag intervals and also on the absolute magnitude threshold for the
quasar sample. Results for the value of the exponent $\beta$ are listed in 
Table\,\ref{tab:sf}. The uncertainties given there are the 
rms errors in the regression only. The total errors are larger and are expected to be 
dominated by the uncertainties in the correction for measurement errors, which
may be underestimated here especially in the u band. (An underestimation lowers
the SF slope.) For $M_{\rm i} < -22$, the mean value of $\beta$,
averaged over the high-throughput SDSS g,r, and i filter bands,  
is $\beta = 0.33\pm0.01$  for $\tau_{\rm r} > 10$ days 
and $0.31\pm0.03$ for $\tau_{\rm r} > 100$ days, respectively. 
The latter value is probably more representative because the SF is less 
sensitive to the correction for measurement errors at longer lags.  

A summary of SF slopes found in the literature is given in Table\,\ref{tab:sf_lit}.    
The values for $\beta$ scatter far more than expected from the individual errors.
However, one has to take into account that the results are based not only on 
different quasar samples but also on different definitions of the SF (last column; see
Sect.\,\ref{sf}), different methods of averaging, different corrections of 
measurement errors, and different time baselines. Therefore, Table\,\ref{tab:sf_lit} 
also lists the definition and the maximum time-lag $\tau_{\rm max}$
(where ``of'' and ``rf'' refer to observer frame and rest-frame, respectively).
The simple average of the previous results based on the SF$^{\rm (A)}$,
except for the Seyfert value from Hawkins (\cite{Hawkins02}),
yields $0.30\pm0.07$ ($0.34\pm0.11$ for all data) which is close to the value 
from the present study. 
Frequently-quoted model predictions are $\beta = 0.83\pm0.08, 0.44\pm0.03,$ and
$0.25\pm0.03$ for random superpositions of supernovae in the starburst model, 
instabilities in the accretion disk, and microlensing due to compact lenses in the 
foreground, respectively (Kawaguchi et al. \cite{Kawaguchi98}; Hawkins \cite{Hawkins02}).
A tendency of the SF slope to increase with wavelength in 
the observer frame might be indicated in Table\,\ref{tab:sf} 
(see also Schmidt et al. \cite{Schmidt10}; their Fig.\,3).

The amplitudes of the SFs shown in Fig.\,\ref{fig:sf} generally agree  with 
findings in the literature listed in Table\,\ref{tab:sf_lit}. De Vries et al. 
(\cite{deVries05}; their figure \,18) 
present a noise-corrected SF for combined g and r band as a function of
time-lag in the observer frame. The last data point in our Fig.\,\ref{fig:sf}
corresponds to time-lags of $\sim 7$\,yr where the de Vries SF predicts log\,$V = -1.08$,
in perfect agreement with log\,$V = -1.0\pm0.1$ for g and r in our study.
De Vries et al. also show (their figure \,9) that the time-lag shifted SF 
from Hawkins (\cite{Hawkins02}) agrees with their data. The SF derived by
Hook et al. (\cite{Hook94}) corresponds to a lower amplitude of
log\,$V = -1.54$ at $\tau \sim 7$\,yr (observer frame). The maximum lag of
the data from Bauer et al. (\cite{Bauer09}) is $\sim 10^3$\,d (rest-frame),
but the amplitudes are not clearly defined at the longest time-lags. For
log\,$\langle \tau_{\rm r}\mbox(d)\rangle = 2.5$, Bauer et al. give log\,$V = -1.6$
without specifying the filter band, compared with -1.3 or -1.4 for g and r in 
the present paper. However, Bauer et al. emphasize that 
``the exact values for $V$ should not be taken seriously \ldots because the 
data are arbitrarily normalized''. From the SFs presented by Wilhite et al.
(\cite{Wilhite08}; see also Rengstorf \cite{Rengstorf06}) we derive 
log\,$V = (-0.99, -1.17, -1.36, -1.44)$ for (u,g,r,i) at their maximum time-lags 
log\,$\langle \tau_{\rm r}(\mbox{d}) \rangle = 2.7$, in good agreement with 
the values $(-1.0, -1.15, -1.28, -1.35)$ found here for the same $\tau_{\rm r}$.
The variability amplitudes derived from the SFs shown by Vanden Berk et al. 
(\cite{VandenBerk04}) are almost the same as in Wilhite et al. (\cite{Wilhite08}).
Kawaguchi  et al. (\cite{Kawaguchi98}) discuss the SF of only one quasar, 
hence their derived variability amplitude is not necessarily representative. 
We emphasize that these results are derived from SFs
defined in different ways and computed from different quasar samples, 
that cover different ranges of time-lags, and are related to different frames 
(rest or observer).

\subsection{Variability as a function of luminosity and redshift}\label{var_l_z}

\begin{figure*}[bhtp]   
\vspace{13.1cm}
\includegraphics[bb=590 1 1 845,scale=0.63,angle=-90,clip]{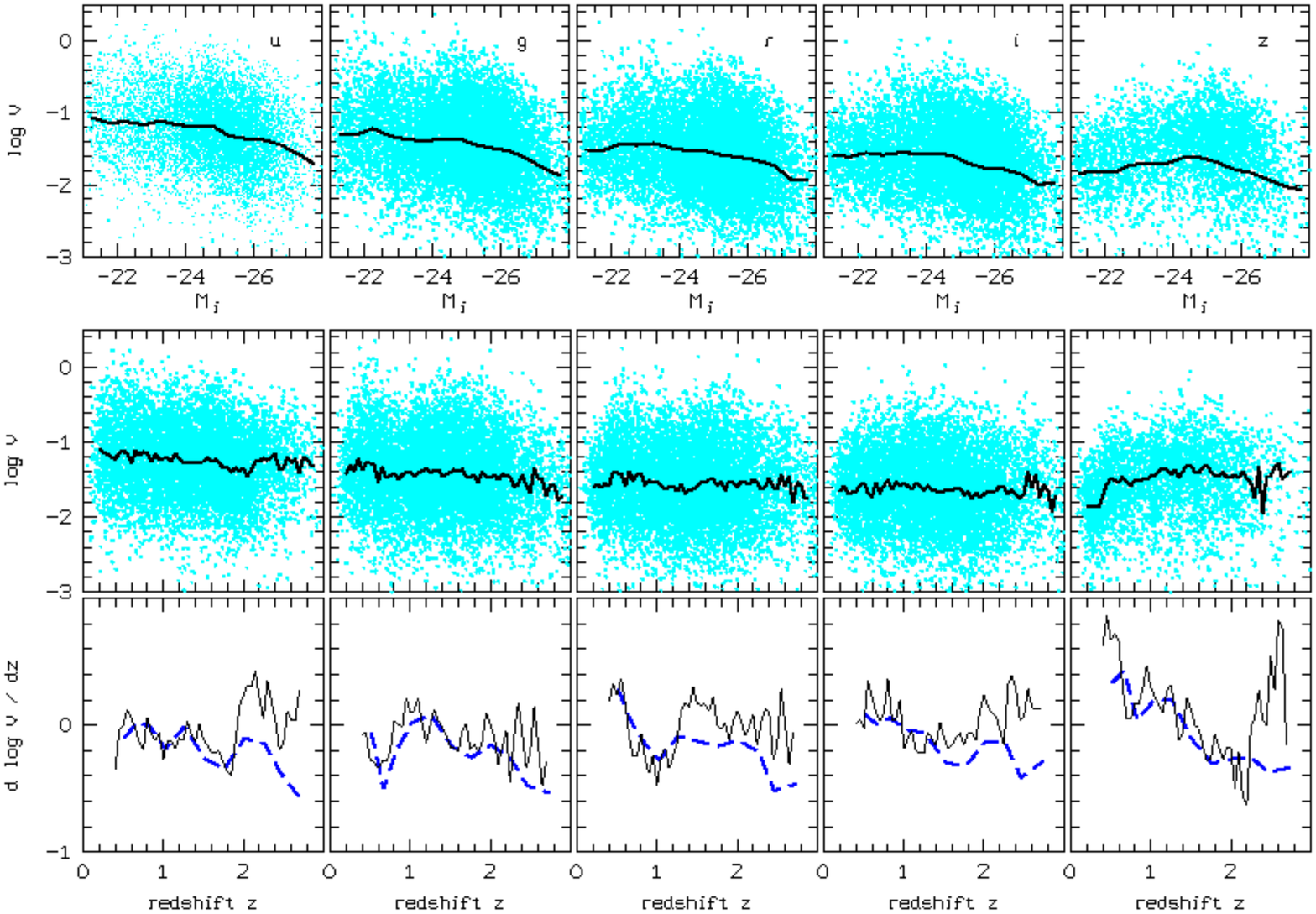}
\caption{
Variability-absolute magnitude relations ({\it top}) and
variability-redshift relations ({\it middle}) for the five SDSS bands.  
The bottom line compares the measured $z$ gradient of variability (solid)
with the $z$ gradient expected from the dependence of $V$ on 
$M_{\rm i}(z)$ (dashed). (Note that the $M_{\rm i}$ are $K$-corrected 
in the rest-frame, yet the $V_{\rm k}$ refer to photometric bands defined
in the observer frame.)
} 
\label{fig:var_l_z}
\end{figure*}

The sample-averaged variability measured in a given photometric passband 
is expected to change with redshift because of the dependence of variability
on absolute magnitude and intrinsic wavelength. In addition, there might
be an explicit dependence of variability on redshift, i.e.,
$V  = V[M_{\rm i}(z), \lambda(z); z]$ and
\begin{equation}\label{eqn:dV_dz}
\frac{\mbox{d} \log V}{\mbox{d}\,z} 
= \frac{\partial \log V}{\partial M_{\rm i}} \frac{\mbox{d}\,M_{\rm i}}{\mbox{d}\,z}
+ \frac{\partial \log V}{\partial \lambda} \frac{\mbox{d}\, \lambda}{\mbox{d}\,z}
+ \frac{\partial \log V}{\partial z},
\end{equation}
where $\lambda$ is the rest-frame wavelength.

Figure\,\ref{fig:var_l_z} shows the distribution of the variable quasars
($\chi^2 > 3$) in the log\,$V$-$M_{\rm i}$ diagrams (top) and 
the log\,$V$-$z$ diagrams (middle), respectively, for the five bands.
We overplot the values of $\log \langle V\rangle$, where 
$\langle\rangle$ denotes the median within bins of the width 
0.5\,mag for $M_{\rm i}$ and 0.05 for $z$, respectively. For luminous quasars 
(here: $M_{\rm i} \la -25$), the average variability decreases with luminosity.
The relatively small variability of low-luminosity AGNs in the z band is 
expected because of the contribution of the host galaxy.
The correlation with redshift is less pronounced but a slight tendency
for larger variability at lower redshifts is apparent for the u and g band.

The solid polygon curves in the bottom row of Fig.\,\ref{fig:var_l_z} 
are the gradients $\mbox{d} \log V/\mbox{d}\,z$ (i.e., the left-hand side of 
Eq.\,\ref{eqn:dV_dz}) derived from the median relations shown in the middle row. 
The dashed lines represent the fraction corresponding to the $M_{\rm i}$-$z$ relation 
(i.e., the first term on the right-hand side of Eq.\,\ref{eqn:dV_dz}) 
computed from the $\log \langle V\rangle$-$M_{\rm i}$ relation from the top line and the 
$\langle M_{\rm i}\rangle$-$z$ relation from Fig.\,\ref{fig:z-Mi}.
In principle, the difference between the two curves can be used
to derive the wavelength dependence of variability provided that there 
is no significant explicit redshift dependence of the variability.  
In particular, that the solid curves have generally higher values than the dashed 
curves at higher $z$ indicates that $V$ increase towards lower intrinsic
wavelengths.  However, the $V$-$\lambda$ relation derived in that way is much
too noisy to provide useful information. The observed $V$-$z$
relation is dominated by the $V$-$M_{\rm i}(z)$ relation as 
indicated by the similarity of the two curves.
A more effective approach to the $V$-$\lambda$ relation will be 
discussed in Sect.\,\ref{sec:var_lambda} below.

To search for an explicit $z$ dependence, we compare 
the variability measured at different $z$ but the same intrinsic 
wavelengths for quasars of the same absolute magnitudes.
This can be achieved with the present sample thanks to the
large number of quasars and the simultaneous measurements in five bands. 
We select the variable quasars in the narrow absolute magnitude interval
$M_{\rm i} = -25\pm0.5$, where the coverage in the redshift space
is good for $z \sim 0.5\ldots 2.2$. For each rest-frame wavelength 
$\lambda \sim 1000\ldots 4000$\,\AA, we computed the redshift intervals 
$z_{\rm k}\pm \Delta z_{\rm k}/2$, where $\lambda$ is shifted into the 
u,g,r, and i passbands, respectively, i.e.,
$z_{\rm k} = \overline{\lambda}_{\rm k}/\lambda-1$ and
$\Delta z_{\rm k} = \Delta \lambda_{\rm k}/\lambda$ where
$\overline{\lambda}_{\rm k}$ is the mean wavelength 
and $\Delta \lambda_{\rm k}$ the width of the corresponding band
in the observer frame  ($k=1\ldots4$).

\begin{figure}[bhtp]   
\vspace{6.5cm}
\includegraphics[bb=576 22 20 818,scale=0.32,angle=-90,clip]{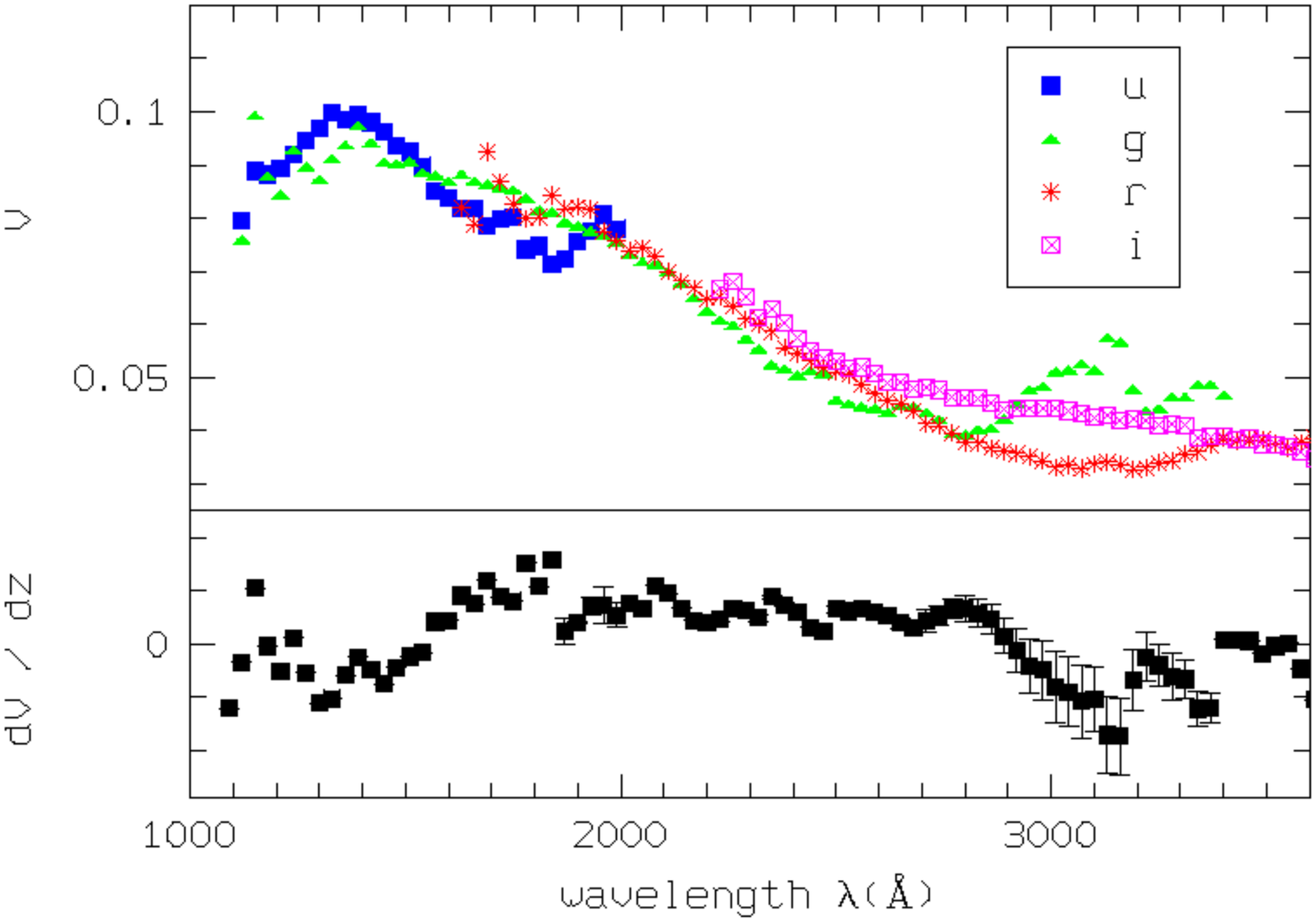}
\caption{
Variability as a function of intrinsic wavelength (rest-frame) ({\it top}) and  
gradient d\,$V/$d\,$z$ ({\it bottom}) for quasars with
$-24.5 > M_{\rm i} >-25.5$ (same scale for both parts of the diagram).
} 
\label{fig:sf_z}
\end{figure}

The top panel of Fig.\,\ref{fig:sf_z} shows the mean variability 
in the redshift intervals around $z_{\rm k} (\lambda)$
as a function of intrinsic wavelength $\lambda$. At fixed $\lambda$,
the band $k$ corresponds to the redshift $z_{\rm k}(\lambda)$. At 
wavelengths where there is an overlap of measurements in at least two bands,  
the comparison of the curves provides information on $V(z)$. 
With a mean value of  
$\langle \mbox{d} V/\mbox{d} z \rangle = 0.001\pm0.007$ 
for $1000\,\mbox{\AA} < \lambda < 3500$\,\AA, 
it is clearly seen (bottom panel) that there is no significant 
explicit variation in $V$ with $z$ for medium-luminosity 
quasars between $z\sim 0.5\ldots 2$. In addition, an increase in 
variability towards shorter wavelengths is clearly apparent in all 
bands (top panel). We note however that the absolute values of $V(\lambda)$ 
are not representative for the whole sample because of the $V$-$M_{\rm i}$
relation.

\subsection{Variability as a function of intrinsic wavelength}\label{sec:var_lambda}

As outlined in Sect.\,\ref{spec_var}, variability ratios
$Q_{\rm k,l}$ were computed for each quasar to measure how variability 
changes from band to band, thus with wavelength. 
The fundamental advantage is that the $Q_{\rm k,l}$ are expected to be
much less (or not at all) contaminated by a dependence of variability
on luminosity (Sesar et al. \cite{Sesar07}). The approach is based 
on the assumption that quasar variability is dominated by the same
processes at high and low  redshifts. We argue that this assumption 
is supported by our finding that there is no indication of an explicit 
dependence of variability on redshift in our sample. The variability ratios 
refer to the same intrinsic time-lags of 300 to 600 days for all 
redshifts. 

In the following, we consider only AGNs with 
$M_{\rm i} <-21$ and $z = 0.2\ldots 3$ because of the small number 
per bin outside this redshift interval. As in the previous 
subsections, we remove the relatively small fraction of unusual BAL quasars, 
core-dominated radio-loud quasars, and the few objects with uncertain 
redshifts. The final number of quasars is large enough to bin the data into 
small $z$ intervals of width $\Delta z = 0.05$. In each of the 56 
$z$ bins, the sample-averaged ratio $\langle Q_{\rm k,l}(z)\rangle$ 
carries information about the variability at five different intrinsic 
wavelengths $\lambda_{\rm k}$. The ensemble variability is thus measured at 
$56\times 5 = 280$ different intrinsic wavelengths. 
Across the wavelength range $\lambda \sim 800 \ldots 8500$\AA,
the formal resolution of the intrinsic variability spectrum 
is therefore (8500$-$800)\AA $/280 \sim 28$\AA. The spectral resolution 
can be improved by a factor $\sim 2$ by choosing narrower $z$ bins. 
 
\begin{figure*}[hpbt]
\begin{tabbing}
\includegraphics[bb=100 426 520 755,scale=0.335,clip]{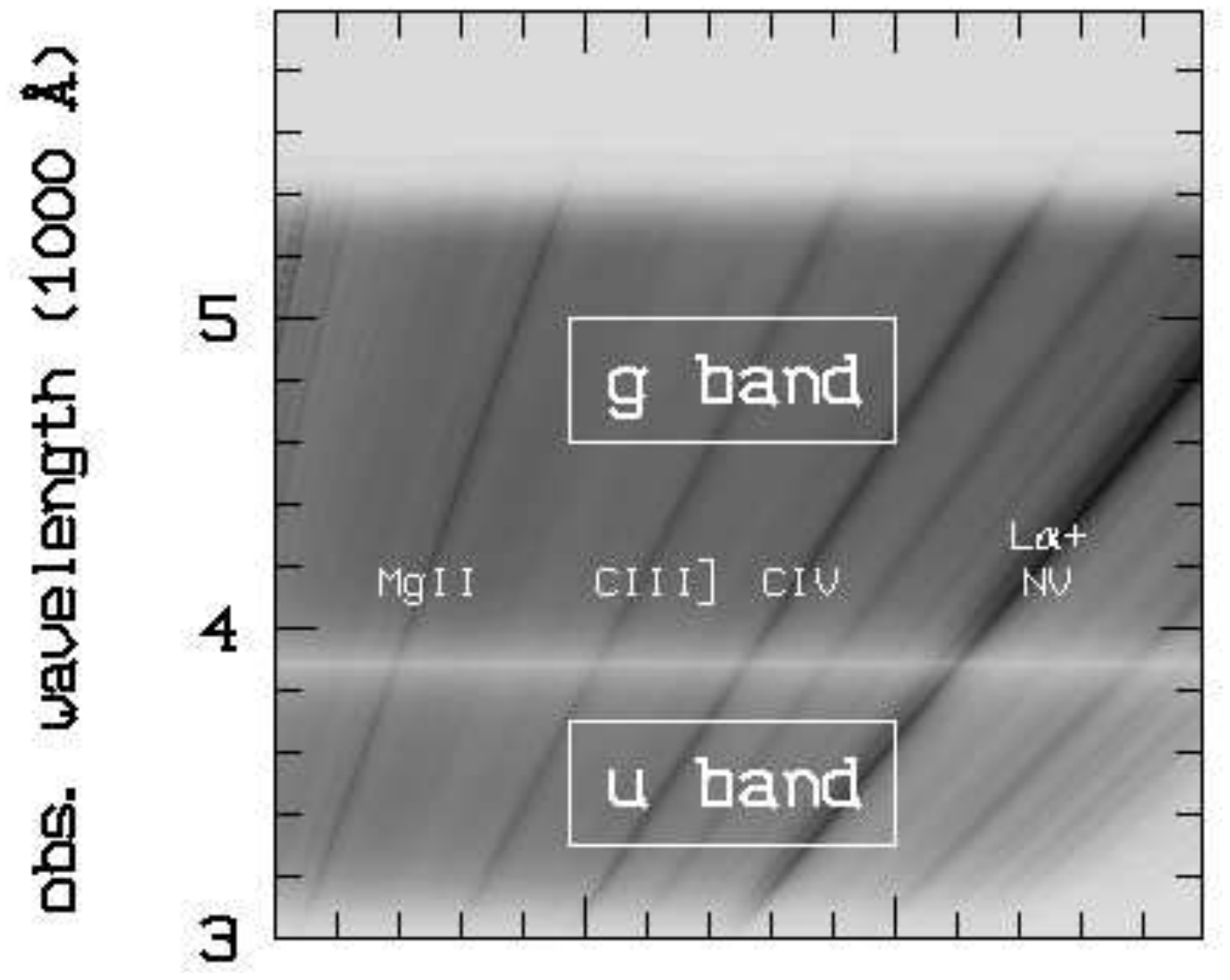}\hfill \=
\includegraphics[bb=165 426 520 755,scale=0.335,clip]{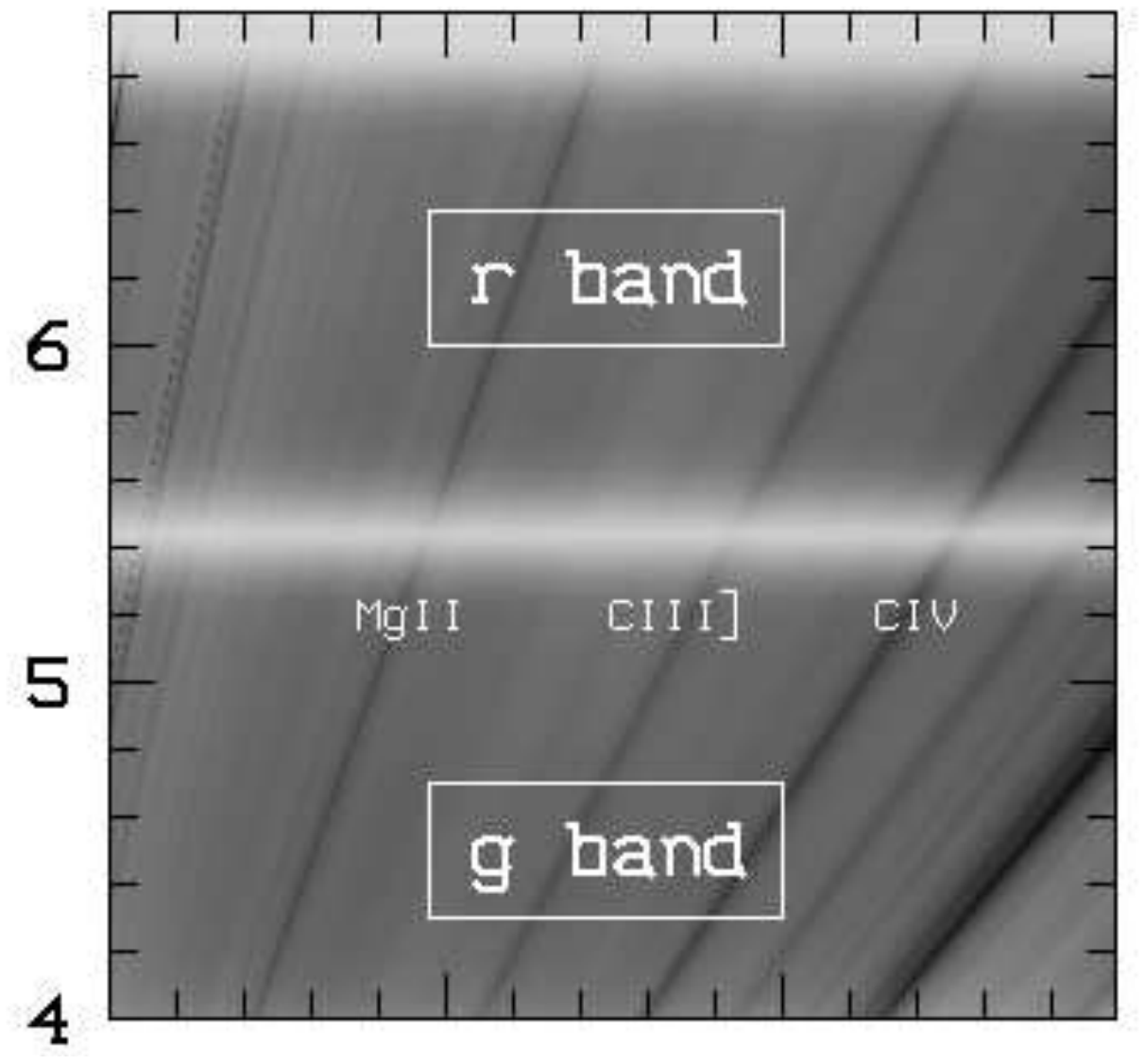}\hfill \= 
\includegraphics[bb=165 426 520 755,scale=0.335,clip]{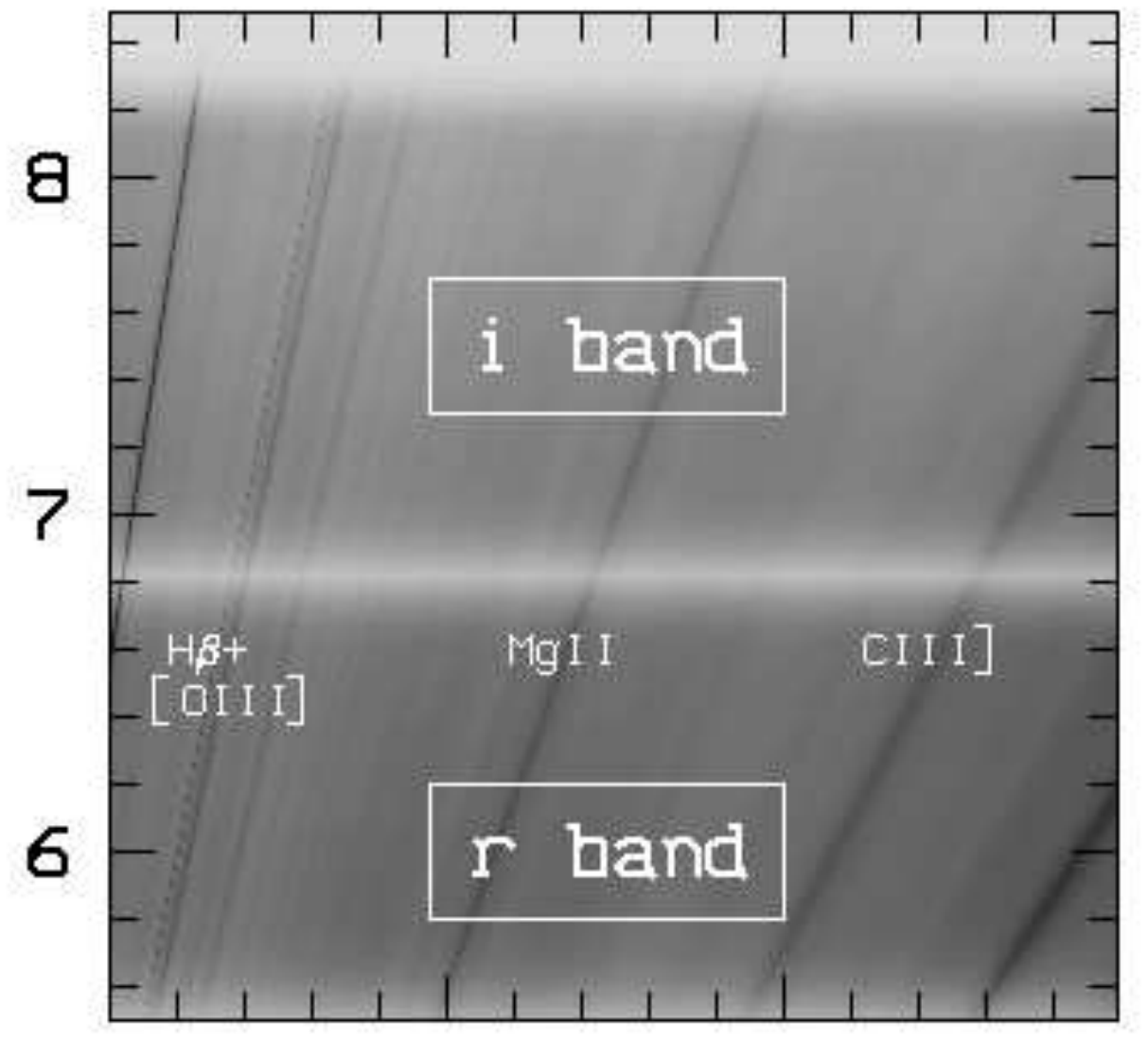}\hfill \=
\includegraphics[bb=165 426 520 755,scale=0.335,clip]{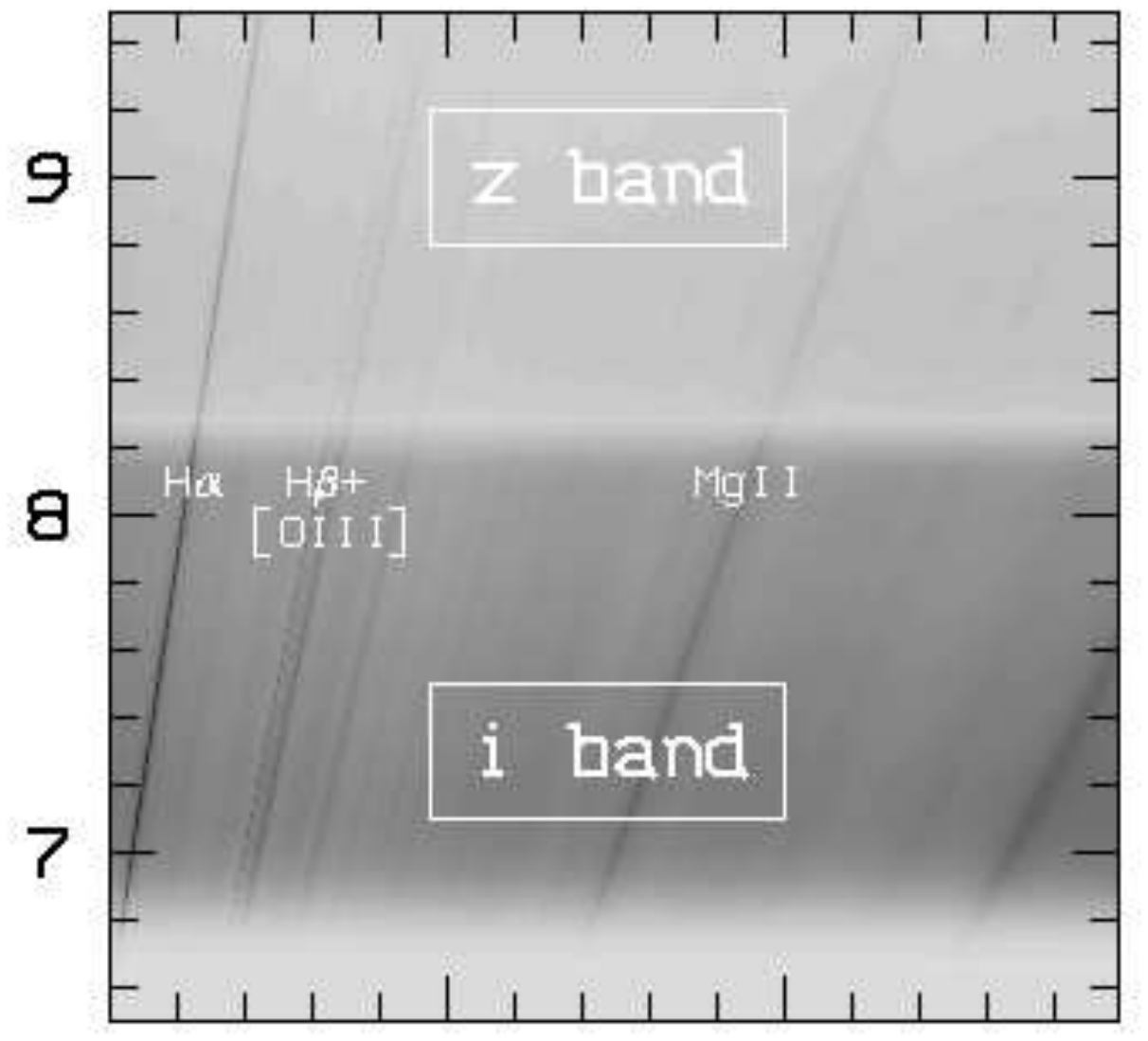}\hfill \\
\end{tabbing}
\vspace{11.5cm}
\includegraphics[bb=590 1 1 842,scale=0.6,angle=-90,clip]{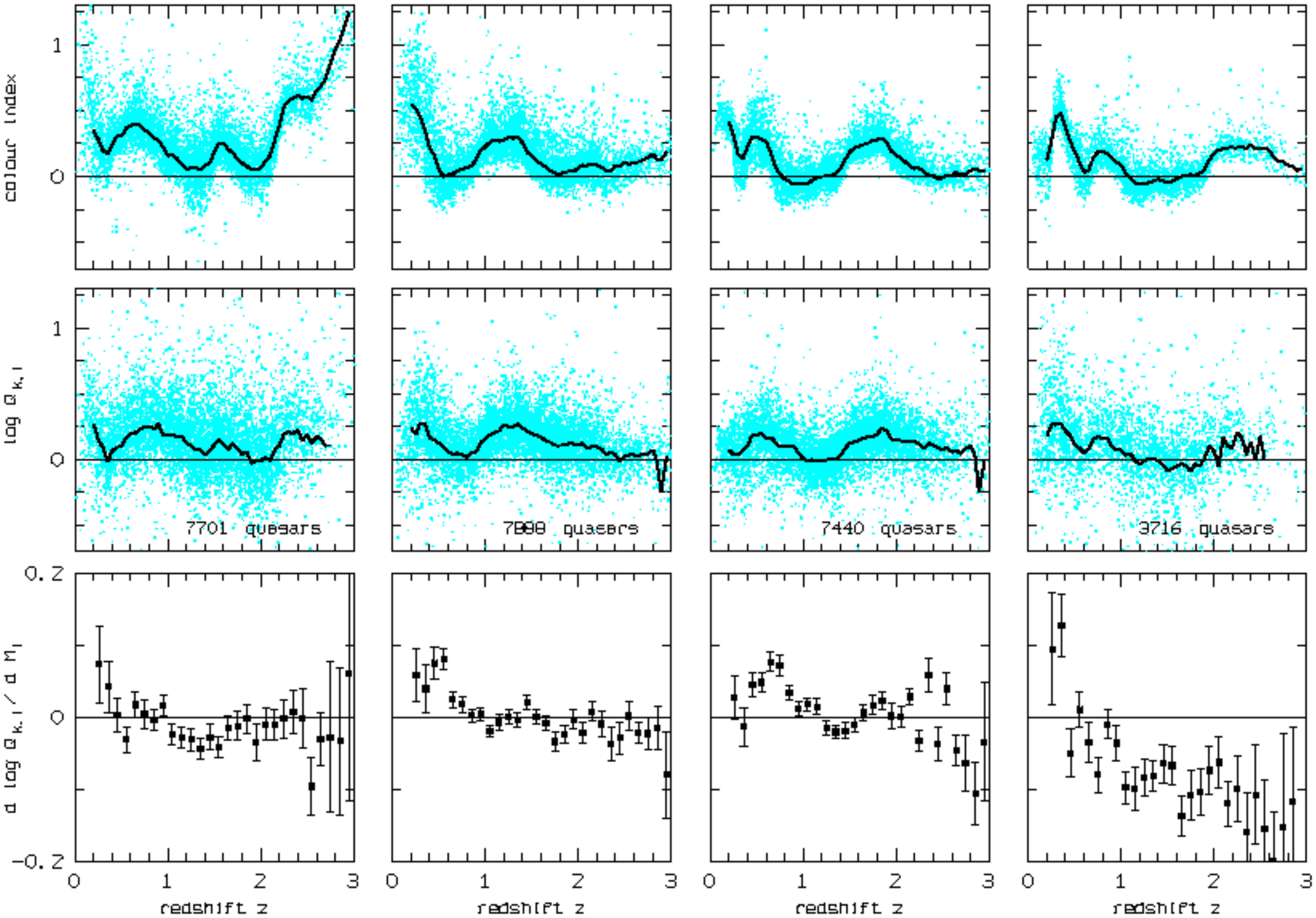}\hfill 
\caption{
{\it Top:} Spectral coverage of the quasar composite spectrum by two adjacent 
photometric bands $k,l$ ($k=$ u,g,v,r and $l=$ g,v,r,i,z) as a function of redshift $z$.
The next rows (top to bottom) show the corresponding colour indices $m_{\rm k}-m_{\rm l}$,
the variability ratios $Q_{\rm k,l}$, and the gradients $d\,\log Q_{\rm k,l} /
d\,M_{\rm i}$, respectively. Polygons: medians of colour index and log\,$Q_{\rm k,l}$,
respectively.
} 
\label{fig:sf_ratios}
\end{figure*}

The results are shown in the third row of Fig.\,\ref{fig:sf_ratios}.
The dots are the $Q_{\rm k,l}$ per quasar on a logarithmic scale with
$k$ and $l$ referring to the SDSS bands ($\lambda_{\rm k} < \lambda_{\rm l}$)
shown in the first row of the same column. The curves are the
log\,$\langle Q_{\rm k,l}\rangle$ - $z$ relations where $\langle\rangle$
denotes the median in $z$ bins of the widths 0.05.
Only quasars with significant variability ($\chi^2 > 3$; Sect.\,\ref{data})
in both the k band and the l band were included. The number of involved quasars 
is given in each panel. In the high-throughput bands, 
more than 7000 variable quasars are available. 
For most redshifts variability is, on average, stronger in the shorter wavelength
bands than at longer wavelengths. 
But it is also obvious from Fig.\,\ref{fig:sf_ratios} that 
the variability ratios display a complicated, nearly undulating dependence on $z$.

The $z$-dependence of the variability ratios can be qualitatively understood
by comparing with the second row of Fig.\,\ref{fig:sf_ratios} where 
the colour indices, $m_{\rm k}-m_{\rm l}$, are shown as a function of redshift. 
The variability ratios and the colour indices have a very similar 
dependence on redshift, though the scatter is clearly larger for the former. 
This similarity was first noted by Sesar et al. (\cite{Sesar07}) for 
the g and r band in the redshift interval $1.0\ldots 1.6$ and is shown here 
to also exist for the other bands and across a broader $z$ range.   
The structure in the colour index-redshift diagrams is known to be caused
by the strong emission lines. To illustrate this
effect, we show in the top row of Fig.\,\ref{fig:sf_ratios} the quasar 
composite spectrum in the observer frame (vertical direction) as a function 
of $z$ (horizontal direction). The spectrum has been multiplied with the 
transmission curves of the corresponding SDSS bands. The most prominent
emission lines are labelled.  The colour index $\Delta m_{\rm k,l}$ decreases
when a strong line enters the band $k$ and increases at redshifts where the 
line leaves the band $k$ and enters the band $l$.  The effect is most 
clearly indicated by the \ion{Mg}{ii}\,$\lambda 2800$\AA\ line where 
there are no other strong lines around. The line dominates the u,g,r,i,z 
band at $z\sim 0.3,0.7,1.2,1.7,2.2$, respectively.
$u-g$ is consequently quite blue at $z\sim 0.3$ but redder at 0.7,
$g-r$ is blue at 0.7 but red at 1.2 etc..

The variability ratios vary with $z$ in the opposite way to the 
flux ratios. When the flux ratio $f_{\rm k}/f_{\rm l}$ increases because a 
strong line enters the band $k$, the variability ratio decreases.
This can be explained only by the assumption that lines are significantly 
less variable than the continuum at the same wavelengths. 
Strong emission lines ``dilute'' the variability in the corresponding band. 
Hence, the $\langle Q_{\rm k,l}\rangle$-$z$ 
relation contains information on the wavelength
dependence of the quasar variability and on the relative 
fraction of variability in the continuum and in the lines. In the 
next section, we use a simple parametrized model to quantify 
this relation. There is no hint of a correlation between the deviations of 
the flux ratios and the variability ratios from their median
relations. 

The bottom line of Fig.\,\ref{fig:sf_ratios} shows that,
at a given redshift, the variability ratio $Q_{\rm k,l}$ also depends 
on luminosity. Interestingly, the gradient 
${\rm d}\,\log\,Q_{\rm k,l}/{\rm d}\,M_{\rm i}$ varies in a way with
$z$ that resembles the variation in the variability ratios itself.

%
%
\section{Discussion}\label{interpretation}
%
%

\subsection{Numerical simulations}\label{sim}

\begin{figure*}[bhtp]   
\vspace{17.1cm}
\includegraphics[bb=587 100 10 740,scale=0.82,angle=-90,clip]{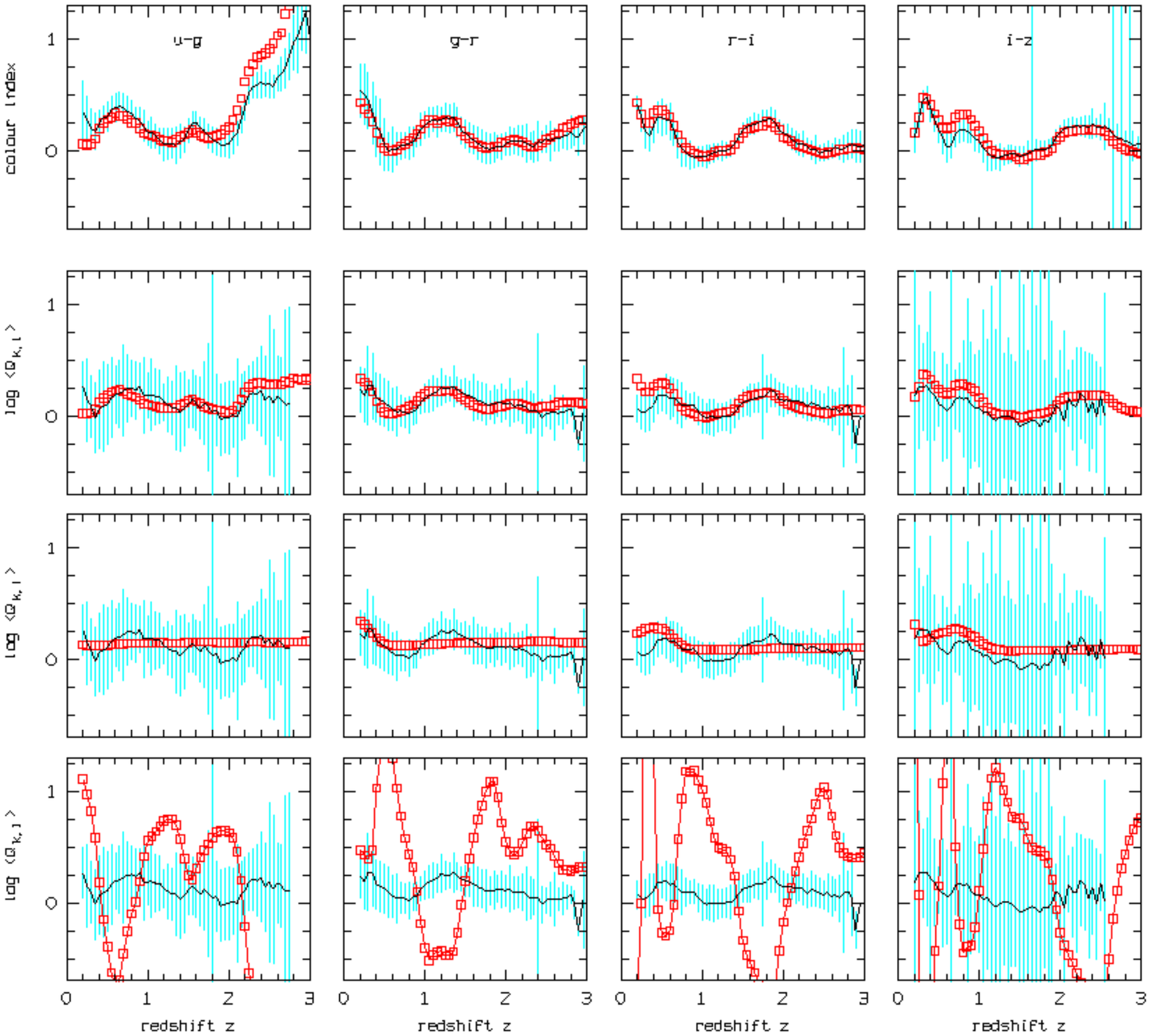}\hfill 
\caption{
Comparison of simulations (open symbols) and observations (thin curves and $1\sigma$ 
deviations): colour indices ({\it top}) and variability ratios 
as functions of redshift for the best-fit model with with small line variability 
($\kappa=0.11$; {\it second row}),
the same model with large line variability ($\kappa=1$; {\it third row}), 
and for a model where only the line flux is variable ({\it bottom}),
respectively. 
} 
\label{fig:sf_ratios_sim}
\end{figure*}

To interpret the results from Fig.\,\ref{fig:sf_ratios} quantitatively, 
we performed Monte Carlo simulations of variable quasar spectra. As a first step, 
the composite spectrum was decomposed into three components: the continuum 
$f_{\rm cont}$, the emission line portion $f_{\rm lines}$, and the host galaxy 
contamination $f_{\rm host}$.\footnote{The fluxes are always related to wavelength intervals,
i.e., $f_\lambda$.} We assumed that the continuum is 
the underlying variable source, the lines are much less variable, and the 
contribution of the host does not vary.

A simple method was used for the decomposition. The continuum $f_{\rm cont}$ was 
described by the power law fitted to the 
composite between Ly\,$\alpha$ and 4000\,\AA, $f_{\rm cont} \propto \lambda^{-0.48}$ 
(Sect.\,\ref{sample}). The power law was extrapolated down to 800\,\AA\ and a 
wavelength-dependent correction factor $c_{\rm H}(\lambda)$ for the heavy Ly\,$\alpha$ forest 
absorption was derived shortward of Ly\,$\alpha$. The product $c_{\rm H}\cdot f_{\rm cont}$ was
subtracted from the composite and the result shortward of 4000\,\AA\ was identified
with the UV emission-line portion. The result longward of 4000\,\AA\ is the 
superimposition of the optical emission lines with the host spectrum. The host contribution 
was approximated by fitting by eye a smooth function to the stellar continuum. After
parametrizing the variability as a function of wavelength, this rough 
decomposition enabled us to simulate variable spectra and analyse at least the 
general trends.

The variability was parametrized by assuming Gaussian-distributed magnitude fluctuations, 
$\Delta m$, (see Sect.\,\ref{data}). The standard deviation $\sigma_{\rm \Delta m}(\lambda)$
is described by a power-law dependence on the intrinsic wavelength $\lambda$.
Hence, the variable portion of the spectrum was represented by
$f_{\rm var}(\lambda) = \overline{f}_{\rm var}(\lambda)\cdot 10^{-0.4\,\Delta m}$ 
where $\Delta m \equiv m-\overline{m} = -2.5\,\log f_{\rm var}/\overline{f}$
and $\overline{f}_{\rm var}$ was the variable portion of the composite spectrum.
The exponent was computed as $\Delta m = \Delta m_{\rm norm} \sigma(\lambda)$
where $\Delta m_{\rm norm}$ is a random number for each single realization of a 
variable spectrum. For $n$ realizations of the same spectrum, $\Delta m_{\rm norm}$
follows a centred and normalized Gaussian distribution. 

The aim of the simulation was to compute the variability ratios $Q_{\rm k,l}$ as a
function of redshift to estimate the parameters of the power law  as well as
the line variability fraction $\kappa$ from the comparison of the simulated with the
observed data.  We adopt $\overline{f}_{\rm var}(\lambda) 
= c_{\rm H}(\lambda)\cdot f_{\rm cont}(\lambda) + \kappa\cdot f_{\rm lines}(\lambda)$
with $\kappa = 0\ldots 1$. Hence, a realization of a quasar spectrum is given by
$f(\lambda) = f_{\rm var}(\lambda)+(1-\kappa)\cdot f_{\rm lines}(\lambda) + f_{\rm host}(\lambda)$.
The simulations cover the same redshift range as the observations.
After transforming into the observer frame, the ``observed'' spectrum is multiplied
by the response functions of the five SDSS filter bands and the total flux in each band is
transformed into a magnitude. From these data, colour indices and variability 
ratios are computed. 

A parameter study was performed to find the best-fit model with 20 
realizations of a spectrum per redshift bin. The Mersenne Twister 
MT19937\footnote{http://www.dm.unito.it/~cerruti/primi/primigrandi/twister.html} 
random number generator was used, which passes various tests such as the stringent 
"dieharder" test suite for 
randomness\footnote{http://www.stat.auckland.ac.nz/dsc-2007/abstracts/eddelbuettel06Oct16.pdf}. 
The agreement between the simulated variability ratios $Q_{\rm sim}$
and the observed ratios $Q_{\rm obs}$ is measured by the sum of the
$(Q_{\rm sim} - \langle Q_{\rm obs}\rangle )^2$ 
over all redshift bins, where $\langle \rangle$ 
indicates the average in the redshift bin.
We started with the assumption that the variability in the lines
is negligible ($\kappa = 0$). The best fit is found for 
\begin{equation}
 \sigma_{\rm \Delta m}(\lambda) 
 = 0.06 + 0.07\cdot \Big(\frac{\lambda}{4000\,\mbox{\AA}}\Big)^{-0.28}.
\label{eqn:sigma}    
\end{equation}
The parameter $\kappa$ was subsequently varied; closest agreement was found for 
$\kappa \sim 0.1$.

The simulated variability ratios of our best-fit model are shown in the second row of 
Fig.\,\ref{fig:sf_ratios_sim}. The general agreement between the
observed and the simulated relations is good. Most of the features seen in the 
mean observed relations are closely reproduced by the simulations. Exceptions are
the low-redshift range for the redder bands and the high-redshift range for the 
bluer ($z\ga 2$ for $Q_{\rm u,k}$ and $z\ga 2.7$ for $Q_{\rm g,r}$). 
The former is attributed to the rough modelling of the host contribution
to the spectrum. The latter probably indicates an inconsistency of the 
composite spectrum at shortest wavelengths ($\lambda \la 1100$\,\AA) with the 
photometric data producing also a discrepancy for the colour indices. 
It is expected that a more elaborated method for the spectrum decomposition 
and an individual treatment of the strongest lines (Sect.\,\ref{lines}) 
will provide an improved fit.

The third row of Fig.\,\ref{fig:sf_ratios_sim} shows the results from
the same model with $\kappa=1$. The comparison with model $\kappa=0.1$ 
clearly indicates that most of the structure seen in the $Q_{\rm k,l}$-$z$ 
relations is due to the small variability of the lines. In addition, the 
influence of the host galaxy is evident in the redder bands at low redshifts. 
For $\kappa \ga 0.2$, the structure in the simulated curves is noticeably 
smeared increasingly with increasing $\kappa$. Finally, the structure reverses 
for models where only the line component is variable 
(bottom line of Fig.\,\ref{fig:sf_ratios_sim}).

\subsection{Comparison with Wilhite et al. (\cite{Wilhite05})}\label{Wilhite}

\subsubsection{Colour variability}\label{colour}

On the basis of the analysis of spectra of quasars with repeated SDSS spectroscopy,
Wilhite et al. (\cite{Wilhite05}) discuss the redshift dependence of  
colour differences $\Delta (m_{\rm k} - m_{\rm l}) 
= (m_{\rm k} - m_{\rm l})_{\rm bright} - (m_{\rm k} - m_{\rm l})_{\rm faint}$. 
There is a change in colour that is a function of redshift and
the filter used (their fig.\,14). As pointed out by those authors, the
features in these diagrams are related to the relative lack of variability
in the emission lines. In all colours, quasars appear bluer in the bright phase at almost 
all redshifts. This behaviour is modulated by, with changing  
redshift, strong emission lines being shifted in or out of the passbands. 
When a strong line shifted into the band k is less variable than the continuum, 
the fraction of the line flux to the continuum flux is larger in the faint phase 
than in the bright phase making $(m_{\rm k} - m_{\rm l})_{\rm faint}$
bluer than expected from the power-law continuum and the
colour difference $\Delta (m_{\rm k} - m_{\rm l})$ redder. 
The features in fig.\,14 of Wilhite et al. can be directly compared to those
seen in the third line of our Fig.\,\ref{fig:sf_ratios}. For example, the 
first ``red bump'' in their $\Delta (u-g)$ versus $z$ diagram at $z \sim 0.35$
corresponds to the minimum of $Q_{\rm u,g}$ i.e., where the variability in u is 
low compared to g, and the ``blue dip'' at $z \sim 0.75$ corresponds to a maximum 
of $Q_{\rm u,g}$. In this way, all dips in the four colour-redshift diagrams
in fig.\,14 of Wilhite et al. can be identified with bumps in the $Q_{\rm k,l}$
diagrams in our Figs.\,\ref{fig:sf_ratios} and \ref{fig:sf_ratios_sim}.

\subsubsection{Emission line variability}\label{lines}

Wilhite et al. (\cite{Wilhite05}) estimate the fraction of variability
in the lines of from 20\% to 30\% for \ion{C}{iv}, \ion{C}{iii}], and \ion{Mg}{ii}.
This result differs significantly from the 10\% we find from fitting
the variability ratios. For $\kappa > 0.2$, the structures in the
simulated $Q_{\rm k,l}$-$z$ relations are remarkably lower and the agreement
with the observations becomes worse. However, our parameter $\kappa$ is
an integral quality that is averaged over all lines. 

In this context, we note that even our best-fit model is not 
perfect. In particular, the slight systematic shift towards lower $z$
of the structure related to the \ion{Mg}{ii} line in the simulated curves 
is most likely due to differences in the variability fraction of  \ion{Mg}{ii} and
the lines in the 3000\,\AA\ bump. More detailed simulations with an
individual treatment of the strongest lines will certainly improve the fit and 
provide insight into the relative strengths of the variability of these lines.

\subsubsection{Continuum variability}\label{continuum}

\begin{figure}[bhtp]   
\includegraphics[bb=78 142 515 700,scale=0.6,angle=0,clip]{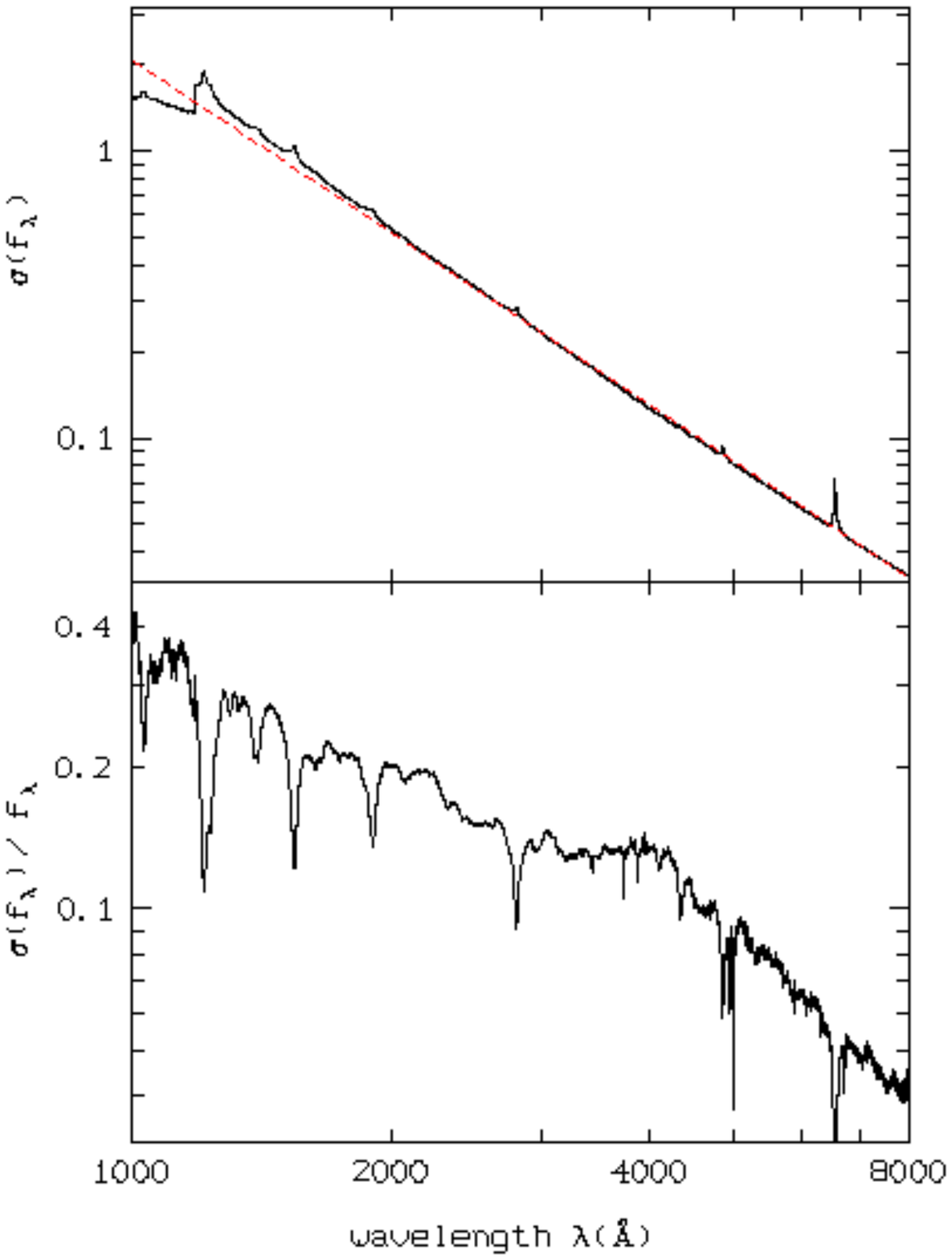}\hfill 
\caption{
{\it Top:} Wavelength dependence of the standard deviation $\sigma$ of the flux density 
$f_\lambda$ for the best-fit model with $\kappa = 0.1$. 
The dashed line corresponds to the power law with slope $-2$. 
{\it Bottom:} Ratio $\sigma (f_\lambda)/\overline{f}_{\lambda}$ where 
$\overline{f}_\lambda$ is the flux density of the composite spectrum normalized at 4000\AA. 
} 
\label{fig:sigma_flux}
\end{figure}

The standard deviation $\sigma$ in the {\it flux} variation as a function of wavelength  
is shown at the top of Fig.\,\ref{fig:sigma_flux} for our best-fit model from
Fig.\,\ref{fig:sf_ratios_sim}. The fluxes in the lines were assumed to vary by 
10\%, hence the lines are visible in this diagram. For $\lambda \ga 2500$\AA, the 
continuum variability is fitted by a power law 
\begin{equation}
\sigma (f_\lambda) \propto \lambda ^{-2}
\end{equation} 
in perfect agreement with the slope of the geometric mean difference spectrum from
Wilhite et al. (\cite{Wilhite05}). This means in particular that the variability
spectrum is steeper (i.e., bluer) than the flux spectrum.

Wilhite et al. (\cite{Wilhite05}) found that the ratio of their arithmetic-mean 
composite difference-spectrum to the arithmetic-mean composite quasar-spectrum 
is almost one for $\lambda \ga 2500$\,\AA\ and the ratio strongly increases
towards shorter wavelengths at $\lambda \la 2500$\,\AA. The ratio of the 
standard deviation in our best-fit model to our arithmetic-mean composite is shown in  
the bottom panel of Fig.\,\ref{fig:sigma_flux}. We confirm that this ratio is  
approximately constant in the wavelength range between the the \ion{Mg}{ii} line and 
$\lambda \sim 4000$\,\AA. However, there is 
a dip shortward of \ion{Mg}{ii} in both our Fig.\,\ref{fig:sigma_flux} (bottom) 
and the corresponding fig.\,13 of Wilhite et al., which leads to the impression 
of an abrupt increase shortward of $\sim 2400$\,\AA. In our model, this dip is
attributed to the very broad conspicuous ``contamination'' feature from 
$\sim 2300$ to 4000\,\AA\,\, (see Vanden Berk et al. \cite{VandenBerk01}). This so-called 
small blue bump (3000\,\AA\ bump) consists of a large number of weak 
\ion{Fe}{ii} emission lines blend together to form a pseudo-continuum  
above the intrinsically emitted continuum, superimposed on the Balmer continuum 
emission (Wills et al. \cite{Wills85}; Vestergaard \& Wilkes \cite{Vestergaard01}).
We argue therefore that the variability increases rather continuously towards short 
wavelengths for $\lambda \la 4000$\,\AA. At longer wavelengths, our ratio 
decreases dramatically, in contrast to that of Wilhite et al.. However, this difference is expected since our sample 
is dominated by low-luminosity AGNs at low redshifts and the contribution from the 
host galaxies ($\lambda \ga 4000$\,\AA) to our composite spectrum is thus 
much larger than in Wilhite et al..

\subsection{Luminosity dependence of variability ratios}\label{luminosities}

The behaviour of the luminosity dependence of the variability ratios as a function 
of redshift (bottom row of Fig.\,\ref{fig:sf_ratios}) can be qualitatively understood
as follows. We assume that $F_{\rm cont, k}$ and $F_{\rm line, k}$ are the total
fluxes of the continuum and the emission lines, respectively, in the band k. 
If the lines are far less variable than the continuum, the fraction of the 
continuum flux $\rho_{\rm k} = F_{\rm cont, k}/(F_{\rm cont, k}+F_{\rm line, k})$ 
is a measure of the strength of variability at a given redshift. 
The variability ratio $Q_{\rm k,l}$ is then expected to scale with the ratio 
$\rho_{\rm k}/\rho_{\rm l}$.
For an average quasar spectrum and at given $z$, the luminosity scales 
with the total flux $F_{\rm tot} = F_{\rm cont} + F_{\rm line}$ in a given band. 
Hence
\begin{displaymath}
G \equiv \frac{\mbox{d} \log\, Q_{\rm k,l}}{\mbox{d} M_{\rm i}} \propto 
-\frac{F_{\rm tot, k}}{Q_{\rm k,l}} \frac{\mbox{d} Q_{\rm k,l}}{\mbox{d} F_{\rm tot, k}}.
\end{displaymath}
For simplicity, we consider the case where there is only one strong emission line
(e.g., \ion{Mg}{ii}) in band k for $z \sim z_1$ and in band l for $z \sim z_2 > z_1$.
Hence
\begin{eqnarray*}
\rho_{\rm k} & = & \Big(\frac{F_{\rm cont}}{F_{\rm tot}}\Big)_{\rm k},
\ \ \ \rho_{\rm l}  =  1,
\ \ \ G_{\rm k} \propto -Q_{\rm k,l}^{-1}\ 
\Big(1+ \frac{\mbox{d} F_{\rm line}}{\mbox{d} F_{\rm cont}}\Big)_{\rm k}^{-1} 
\ \ \ \mbox{for}\ z \sim z_1,\\
\rho_{\rm l} & = & \Big(\frac{F_{\rm cont}}{F_{\rm tot}}\Big)_{\rm l},
\ \ \ \rho_{\rm k}  =  1,
\ \ \ G_{\rm l} \propto -Q_{\rm k,l}\ 
\Big(1+ \frac{\mbox{d} F_{\rm line}}{\mbox{d} F_{\rm cont}}\Big)_{\rm l}^{-1} 
\ \ \  \mbox{for}\ z \sim z_2.\\
\end{eqnarray*}\label{eqn:gradients}

The gradient $G$ depends directly on the variability ratio and, in addition,
on the relationship between the line flux and the continuum flux in the band. 
The equivalent widths of the broad UV emission lines is known for a long time 
(Baldwin \cite{Baldwin77}; Osmer et al. \cite{Osmer94}) to be inversely correlated 
to the luminosity of the underlying continuum in a nonlinear way.
This well-known Baldwin effect appears in two flavours: (i) a global relationship that 
describes fluctuations from object to object and (ii) an intrinsic relationship that
describes how the line flux changes when the continuum of the same object changes
because of variability (Kinney et al. \cite{Kinney90};
Pogge \& Peterson \cite{Pogge92}). In addition, however,
the variation in $G$ with $z$ also reflects the response functions of the
SDSS bands because the ratio of the line flux to the continuum flux in the 
band depends on the throughput at the wavelength of the line. 
To summarize, the $G$-$z$ relation (bottom line of Fig.\,\ref{fig:sf_ratios}) 
reflects the $Q_{\rm k,l}$-$z$ relation, which is, however, modified by both the 
Baldwin effect and the filter response curves.

\subsection{Special quasar types}\label{unusuals}

\begin{figure*}[bhtp]   
\vspace{18.2cm}
\includegraphics[bb=588 118 7 720,scale=0.87,angle=-90,clip]{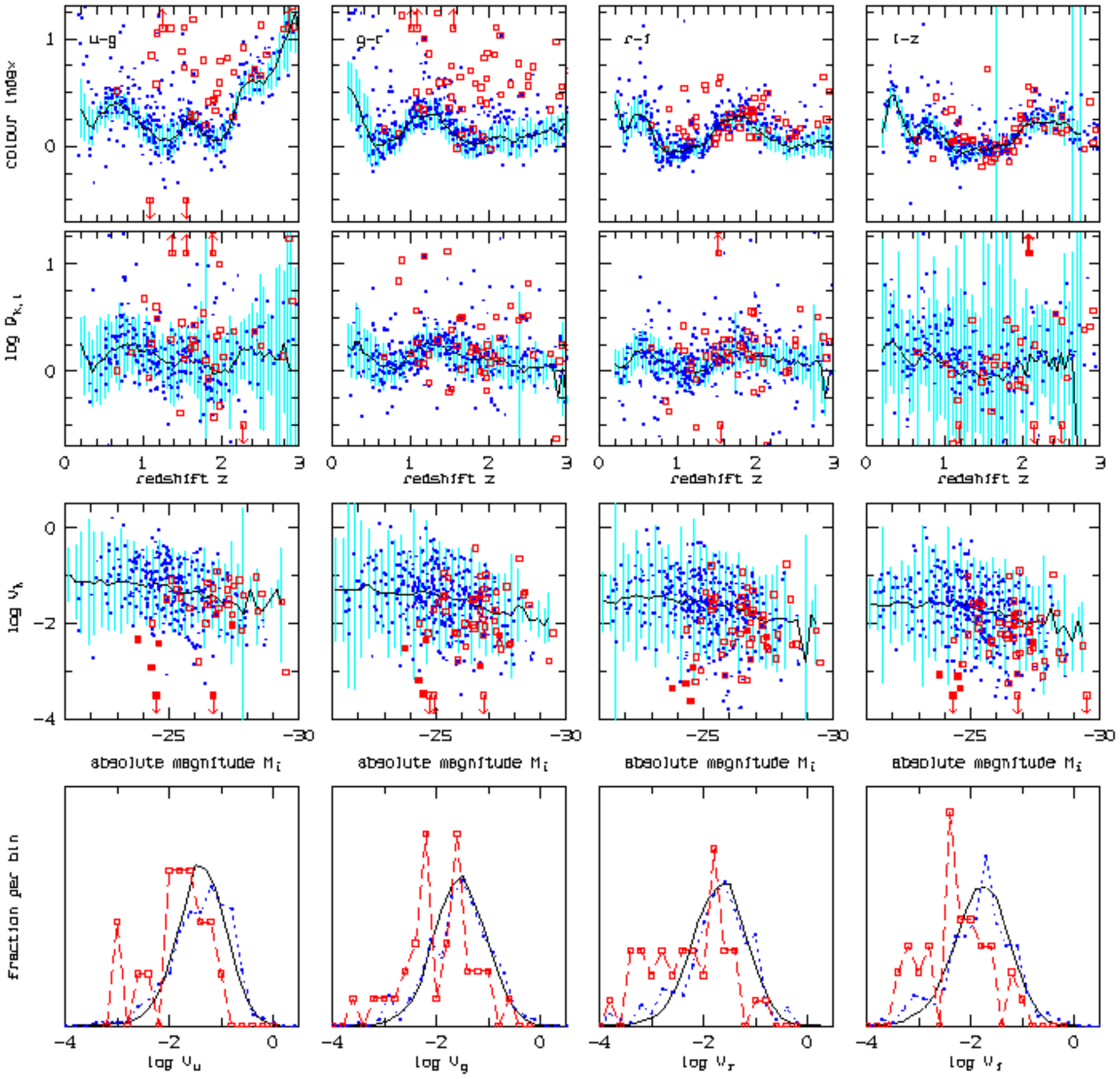}\hfill 
\caption{
Colour indices, $m_{\rm k}-m_{\rm l}$, ({\it top}) and variability ratios, 
$Q_{\rm k,l}$, ({\it second row}) as a function of redshift and variability index, 
$V_{\rm k}$, as a function of absolute magnitude, $M_{\rm i}$, ({\it third row}) 
for radio-loud quasars (dots) and quasars with unusual spectra (squares and arrows),
respectively. Curves and hatched areas as in Fig.\,\ref{fig:sf_ratios_sim}, filled 
squares in the third row: unusual quasars with weak emission lines 
(Fig.\,\ref{fig:unusual_spectra}, left panel). {\it Bottom}: Normalized 
distributions of $\log V_{\rm k}$ for the unusual quasars (open squares and dashed
lines), the radio-loud quasars (dots and dotted lines), and the standard sample 
(solid). 
} 
\label{fig:radio_xBALs}
\end{figure*}

Among the 8744 objects in our final quasar sample, there are 54 quasars with 
unusual spectra and 454 quasars were identified with FIRST radio 
sources. Adopting the criterion 
$R_{\rm i} \equiv \mbox{log}\,(F_{\rm radio}/F_{\rm i}) >1$
(Sect.\,\ref{sample}), we identified 381 radio-loud quasars 
in our sample. In Fig.\,\ref{fig:radio_xBALs}, the optical colours and variability properties 
of both subsamples are compared with those of our standard quasar sample.
As was noted by Richards et al. (\cite{Richards01}) and
Ivezi\'c et al. (\cite{Ivezic02}), the median colours are slightly redder 
and the fraction of red outliers is higher for the radio-loud subsample 
than for the standard sample, where the most extreme outliers are seen 
in the bluer passbands.  A substantial fraction of the radio-loud subsample are 
also outliers in the $Q_{\rm k,l}$-$z$ diagrams. In particular, there
is more than a dozen quasars with low $Q_{\rm k,l}$, i.e., 
$V_{\rm i} > 2 V_{\rm r}$.   

For the subsample of quasars with unusual spectra, the deviation from the standard sample
is much stronger. The colour indices $u-g$ and $g-r$ are much redder than for 
both the standard sample and the radio-loud subsample. These differences 
are again most pronounced for the bluer passbands and vanish for $i-z$. 
This result is not surprising. Both dust absorption and extended and/or overlapping 
BAL troughs in the UV, which are two defining features of this classification, have 
the effect of reddening especially the colour indices of the blue
passbands. In this context, 
the scatter in the variability ratios might be expected to be explained by the 
diversity of the fraction of the emission-line flux in strong and unusual BAL quasars. 

The fraction of radio detections among the unusual quasars is 
17\% (9/54) and the fraction of radio-loud quasars is 11\% (6/54). 
The corresponding values for the whole sample are 5.2\% (454/8744)
and 4.4\% (381/8744). This difference is, of course,  due to SDSS having
targeted FIRST sources for spectroscopy even when they were not
selected as quasar candidates because of their optical colours.   
The three unusual quasars with the highest radio-loudness parameters are 
\object{SDSS\,J000051.56+001202.5} ($R_{\rm i} = 1.89$), 
a high-redshift ($z = 3.859$) BAL quasar with broad Ly$\alpha$,
\ion{Si}{iv}, and \ion{C}{iv} absorption troughs,
\object{SDSS\,J024224.02+010452.5} ($R_{\rm i} = 1.52$),
a medium-redshift ($z = 2.431$) quasar with apparently weak emission 
lines and associated absorption lines, and
\object{SDSS\,J022337.76+003230.6} ($R_{\rm i} = 1.77$),
a weak-line quasar at $z \sim 3.05$. The first two are strongly
variable in the optical, whereas SDSS\,J022337.76+003230.6 exhibits a 
remarkably low level of variability in the u and g bands. 

As a main result of Sects.\,\ref{sim} and \ref{Wilhite}, we have seen that the flux 
in the emission lines is considerably less variable than the underlying continuum.
A relatively low variability $V_{\rm g}$ is thus expected 
at $z\sim3$ where the strong Ly$\alpha$+\ion{N}{v} lines fall into the 
g band (see also Fig.\,\ref{fig:var_l_z}). However, the value of $V_{\rm u}$ 
for SDSS\,J022337.76+003230.6  
is a factor of 20 smaller than the mean value at $z = 3$ and, surprisingly,
SDSS\,J022337.76+003230.6 is a member of the rare class of weak emission-line 
quasars (WLQs; Diamond-Stanic et al. \cite{Diamond09}). WLQs are defined as
quasars with extremely weak or undetectable emission lines of rest-frame 
equivalent width $\mbox{EW(Ly}\alpha+\ion{N}{v}) < 10$\,\AA\  compared to 
$\sim 50\ldots 100$\,\AA\ for typical SDSS quasars (e.g., Fan et al. \cite{Fan99};
Diamond-Stanic et al. \cite{Diamond09}, and references therein). 
For SDSS\,J022337.76+003230.6, Diamond-Stanic et al. measured 
$\mbox{EW(Ly}\alpha+\ion{N}{v}) = 8.6$\,\AA. Quasars with such weak lines might be
expected to exhibit larger variability in the passbands of the lines
than typical quasars. This is clearly not the case for  
SDSS\,J022337.76+003230.6. Altogether the Diamond-Stanic WLQ sample contains four
quasars in S82, three of them being members of our quasar sample where
two have small variability (SDSS\,J022337.76+003230.6 and \object{SDSS\,J005421.42-010921.6})
and the third (\object{SDSS\,J025646.56+003858.3}) has values close to the medians.

If the weak lines are due to the Baldwin effect, WLQs might be expected 
to populate the high-luminosity end of the quasar luminosity 
function, which is obviously not the case (Shemmer et al. \cite{Shemmer09}, and 
references therein). The first interpretation of the nature of WLQs 
involved continuum boosting (Fan et al. \cite{Fan99}) as in local BL Lac 
objects. However, the rest-frame 0.1-5$\,\mu$m spectral energy distributions 
of WLQs does not differ from those of normal quasars and variability, 
polarization, and radio properties differ from those of BL Lacs 
(Diamond-Stanic et al.\cite{Diamond09}). In addition, Shemmer et al. (\cite{Shemmer09})
pointed out that it would be difficult to explain the lack of a large parent 
population of X-ray and radio bright weak-lined sources at high $z$
if WLQs were the long-sought high-redshift BL Lacs. 
Shemmer et al. (\cite{Shemmer06}) argued against the possibility that WLQs are
quasars with continua amplified by microlensing. This interpretation 
is also questioned by the low variability found in the present study based on 
intrinsic timescales of 1 to 2 yr, which are typical timescales for
quasar microlensing.
 
\begin{figure}[bhtp]   
\includegraphics[bb=35 67 560 780,scale=0.5,angle=0,clip]{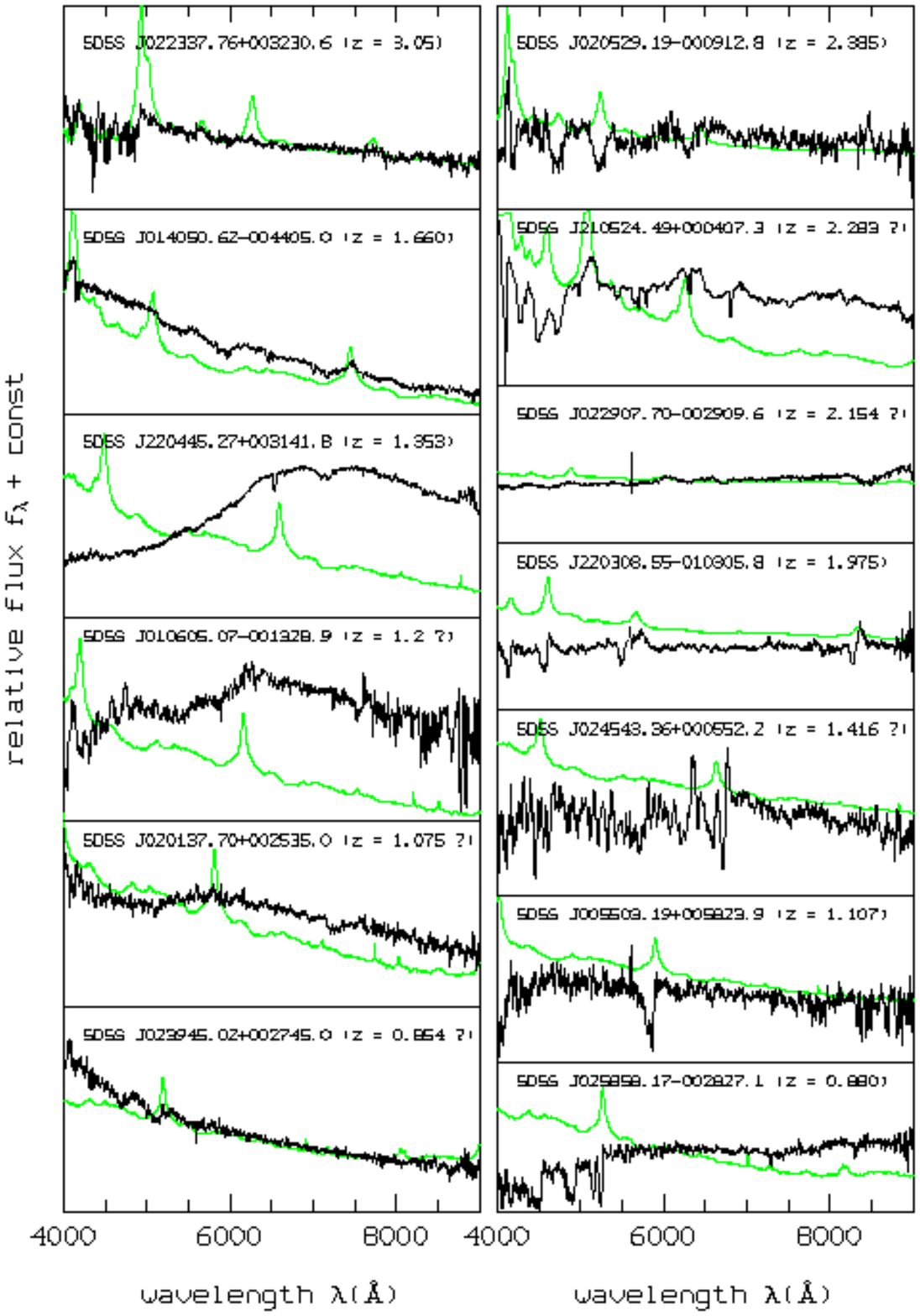}\hfill 
\caption{
{\it Left}: Unusual quasar spectra with weak emission lines.
{\it Right}:  Another 7 unusual quasars with weak variability ({\it right}).
In each panel, the composite spectrum is overplotted as smooth curve
(arbitrarily scaled).   
} 
\label{fig:unusual_spectra}
\end{figure}

Among the quasars classified here as unusual, no additional one matches the 
classical criterion of WLQs (i.e., $\mbox{EW(Ly}\alpha+\ion{N}{v}) < 10$\,\AA). 
We note that our quasar sample is biased against spectra that have only 
a featureless continuum (Sect.\,\ref{sample}). However, we selected (by eye) 
another five quasars for which the emission lines are substantially less pronounced 
than in the composite spectrum but where there is no clear-cut evidence of BAL 
troughs. Their spectra are shown in the left panel of 
Fig.\,\ref{fig:unusual_spectra} along with the WLQ SDSS\,J022337.76+003230.6.
\object{SDSS\,J220445.27+003141.8} ($z = 1.353$) is one of the two ``mysterious objects'' 
from Hall et al. (\cite{Hall02}). The interpretation of 
\object{SDSS\,J010605.07-001328.9} is unclear
but the spectrum resembles the unusual FeLoBAL quasar VPMS\,J134246.25+284027.5 
(Meusinger et al. \cite{Meusinger05}). 
In the $\log V$-$M_{\rm i}$ plots in Fig.\,\ref{fig:radio_xBALs}, these quasars 
are marked as filled squares. All of them vary by less than the general median
dispersion.

Apart from the puzzling WLQs, the variabilities of the unusual subsample 
are systematically smaller than those of the standard sample. The 
bottom row of Fig.\,\ref{fig:radio_xBALs} displays the distributions of $\log V_{\rm k}$
for the unusual quasars, the radio-loud quasars, and the standard sample
in the absolute magnitude range $-27 < M_{\rm i} < -23$. There are no
substantial differences between the radio-loud subsample and the standard sample. 
For the unusual quasars, however, we find that 24.5\% (13/53) of the quasars with 
$M_i < -23$ have $V < 0.01$ in each of the bands u,g,r, and i, 
compared with 6.6\% (483/7334) for the standard sample in the same $M_{\rm i}$
range.  Among these 13 low-variability unusual quasars, there are in particular all 
six WLQ-like objects from the left panel of Fig.\,\ref{fig:unusual_spectra}. 
The spectra of the remaining 7 quasars are 
shown in the right panel of Fig.\,\ref{fig:unusual_spectra},
most of which have strong and/or unusual BAL structures.  
The properties of both
\object{SDSS\,J025858.17-002827.1} and \object{SDSS\,J024543.36+000552.2}
are most likely caused by many low ionisation \ion{Fe}{ii} absorption 
troughs (FeLoBALs). It is remarkable that in all these spectra there is in particular 
a the lack of strong emission lines. 

Why is the variability of the unusual quasars smaller than the median values, even 
though a lack of broad-line emission is expected to result in stronger
variability?
Wilhite et al. (\cite{Wilhite08}) proposed that quasar variability is 
inversely related to the Eddington ratio $L_{\rm bol}/L_{\rm Edd}$, and
Shemmer et al. (\cite{Shemmer09}) suggested that WLQs might be
quasars with high Eddington ratios. Ai et al. (\cite{Ai10}) 
found by comparing broad-line Seyfert\,1 and narrow-line Seyfert\,1 type quasars 
that the variability amplitude correlates with the relative strength of
emission lines and anticorrelates with the Eddington ratio.
If $L_{\rm bol}/L_{\rm Edd}$ is an (at least crude) estimate of the accretion rate, 
both weak emission lines (but strong \ion{Fe}{ii} pseudo-continuum) 
and small variability amplitudes are the consequence of strong accretion. 
An anticorrelation with variability is plausible because,
if the accretion rate is high, the region emitting the continuum observed 
in a given passband is larger and the variability amplitudes 
are consequently smaller than in the case of a low accretion rate 
(Ai et al. \cite{Ai10}). Wilhite et al. (\cite{Wilhite08}) proposed 
that the accretion rate is high and the variability is small 
for optically young quasars, i.e., when they
become visible in the optical. On the other hand, FeLoBAL quasars
have been suggested to represent optically young AGNs expelling a thick 
shroud of gas and dust from the cocoon phase (e.g., Voit et al. 
\cite{Voit93}; Farrah et al. \cite{Farrah10}). In this context, the 
higher fraction of low-variability quasars among the unusual objects 
might be an indication that the unusual subsample contains 
a higher fraction of young quasars. However, the present sample is quite small and 
inhomogeneous. A more detailed investigation of 
the properties of quasars with atypical spectra in S82 is 
necessary.

%
\section{Conclusions}\label{conclusion}
%

We have compiled a sample of nearly 9000 quasars from the SDSS Stripe 82 with redshifts
$z \sim 0.1 \ldots 5$ and absolute magnitudes $M_{\rm i} \sim -20 \ldots -30$
that have multi-epoch photometry in the LMCC (Bramich et al. (\cite{Bramich08}).
The fraction of variable quasars in the sample is found to
be 93\%, 97\%, 93\%, 87\%, and 37\% for the u,g,r,i, and z band,
respectively. The SDSS spectra of all these objects were studied individually.
For the wavelengths range from Ly\,$\alpha$ to $\sim 4000$\,\AA, our arithmetic-mean 
composite spectrum is similar to the composite spectrum from Vanden Berk et al. 
(\cite{VandenBerk01}). At longer wavelengths, our composite is more strongly contaminated by 
stellar radiation from the host galaxies. The continuum is well fitted by a power law with
$\alpha_\nu = -0.48$.

After removal of strong photometric outliers caused by ``bad epochs'' in the 
LMCC lightcurves, first-order structure functions (SFs) were computed and 
used to analyse the ensemble variability of the ``normal'' quasars 
(excluding unusual BAL quasars and the known 
core-dominated radio-loud quasars). The following 
main conclusions are drawn:
\begin{itemize}
\item[1.]
The SF as a function of the time-lag can be fitted by a power law with the index 
$\beta \sim 0.31\pm0.03$ for rest-frame lags $\tau = 100\ldots1000$\,d.
Our data confirm the existence of anti-correlations of variability with both 
luminosity and redshift $z$. The trend with $z$ can be interpreted as a 
consequence of the dependence on both luminosity and intrinsic wavelengths. There
is no indication of an additional redshift dependence. 
\item[2.]
The dependence of variability on rest-frame wavelengths can be established by
studying the ratio of the variability in two different passbands
as predicted by Sesar et al. (\cite{Sesar07}). The redshift 
dependence of the variability ratios, $Q_{\rm k,l}$, resembles that of the 
corresponding colour indices. The $Q_{\rm k,l}$-$z$ relations indicate that
variability is almost always stronger in the blue passband than in the redder 
one. However, the relations are modulated by the appearance of strong 
emission lines in either one band or the other. 
\item[3.]
By means of simple Monte Carlo simulations of spectral variability,
we have shown that the observed $Q_{\rm k,l}$-$z$ relations can be 
described by assuming a variability spectrum $\sigma(f_\lambda) \propto \lambda^{-2}$ 
for the continuum ($\sigma$: standard deviation) and a low percentage of $\sim 10$\% 
of variability in the emission lines. Quasars are mostly bluer in their bright 
phase but this depends on both $z$ and the passbands. The low variability in the 
lines is related to the intrinsic Baldwin effect. These results, based upon
the photometry of $\sim 8000$ quasars, confirm the findings of Wilhite et al. 
(\cite{Wilhite05}) for 315 quasars with repeated SDSS spectroscopy.
\item[4.]
The microlensing scenario for quasar variability is not excluded by the observed 
slope of the SF. On the other hand,  the derived fluctuation spectrum 
is consistent with the assumption that the dominant fraction of the 
optical/UV quasar variability comes from the accretion disk as shown by 
Pereyra et al. (\cite{Pereyra06}). It remains to be seen whether the microlensing
scenario predicts the observed ratio of variability in line flux to
continuum flux. But prior to that, more detailed information on the variability of the 
various line components needs to be extracted from the data.
\item[5.]
For quasars with spectra that deviate significantly from the quasar
composite spectrum, not only the colour-$z$ but also the $Q_{\rm k,l}$-$z$ relations 
differ from those of typical quasars. Unusual quasars with weak or 
undetectable emission lines tend to exhibit smaller fluctuations than normal 
quasars. This trend, obviously opposite to the ``dilution'' of the observed 
variability by the emission line flux, might be an indication that 
a significant fraction of these unusual quasars accrete with high Eddington 
ratios.
\end{itemize}

\begin{acknowledgements}
The thank the referee, Richard Kron, for his helpful
criticism and useful comments. 

Funding for the SDSS and SDSS-II has been provided by 
the Alfred P. Sloan Foundation, the Participating Institutions 
(see below), the National Science Foundation, the National 
Aeronautics and Space Administration, the U.S. Department 
of Energy, the Japanese Monbukagakusho, the Max Planck 
Society, and the Higher Education Funding Council for 
England. The SDSS Web site is http://www.sdss.org/.

The SDSS is managed by the Astrophysical Research 
Consortium (ARC) for the Participating Institutions. 
The Participating Institutions are: the American 
Museum of Natural History, Astrophysical Institute 
Potsdam, University of Basel, University of Cambridge 
(Cambridge University), Case Western Reserve University, 
the University of Chicago, the Fermi National 
Accelerator Laboratory (Fermilab), the Institute 
for Advanced Study, the Japan Participation Group, 
the Johns Hopkins University, the Joint Institute 
for Nuclear Astrophysics, the Kavli Institute for 
Particle Astrophysics and Cosmology, the Korean 
Scientist Group, the Los Alamos National Laboratory, 
the Max-Planck-Institute for Astronomy (MPIA), 
the Max-Planck-Institute for Astrophysics (MPA), 
the New Mexico State University, the Ohio State 
University, the University of Pittsburgh, University 
of Portsmouth, Princeton University, the United 
States Naval Observatory, and the University of 
Washington. 
\end{acknowledgements}


{}
%


\end{document}